\documentclass[%
reprint,
amsmath,amssymb,
aps,
prb,
showkeys
]{revtex4-2}

\usepackage{graphicx}
\usepackage{dcolumn}
\usepackage{bm}
\usepackage{xcolor}
\usepackage{hyperref}
\usepackage{multirow}
\usepackage{array}
\definecolor{red}{rgb}{1.0, 0.01, 0.24}

\date{} 
\begin{document}

\title{\textbf{Disorder-engineered magnetic compensation in trilayered square Ising ferrimagnet: a Monte Carlo study}}

\author{Soham Chandra}
\email{sohamc07@gmail.com}

\affiliation{Department of Physics, Presidency University, 86/1 College Street, Kolkata -700 073, India}
\affiliation{Department of Physics, Brainware University, 398 Ramkrishnapur Road, Barasat, Kolkata-700125, India}%

\begin{abstract}
    Because of its prospective uses in thermomagnetic and spintronic devices, ferrimagnetic multilayer nanostructures displaying compensating behaviour are of great interest. In this study, we examine how the thermomagnetic characteristics of spin-1/2 Ising trilayer ferrimagnets made of coupled square monolayers with ABA and AAB stacking sequences are affected by controlled site dilution. The system is composed of two different types of theoretical atoms, with atoms of the same type (A-A and B-B) exhibiting ferromagnetic interactions, while unlike atoms (A-B) display antiferromagnetic interactions. We examine the effects of randomly added nonmagnetic impurities in the A-layers on the system's magnetisation, susceptibility, specific heat, compensation temperature, and critical temperature using comprehensive Metropolis Monte Carlo simulations. The results reveal that increasing impurity concentration systematically, from 5\% to 45\%, lowers both the compensation and critical temperatures, while preserving the continuous nature of the magnetic phase transition, leading to different equilibrium ferrimagnetic behaviours. More importantly, site dilution is found to induce compensation points in regions of the interaction parameter space where compensation is absent in the pristine systems. Comprehensive phase diagrams in the $(J_{AB}/J_{BB} \times J_{AA}/J_{BB})$ plane are constructed for different impurity concentrations, demonstrating the impurity-driven evolution of compensating and non-compensating phases. We further identify threshold impurity concentrations associated with the emergence of magnetic compensation and establish phenomenological scaling relations connecting the compensation characteristics with the interaction strengths and dilution percentage. The phase area in the Hamiltonian parameter space, \textit{without} compensation, scales according to the relation: $\ln |A(\rho)/\tilde{A}|=ae^{b\rho}$, with configuration dependent $a$, $b$, and $\tilde{A}$ . Suggested mathematical formulas reveal how the threshold impurity concentration is linked to the Hamiltonian parameters. These results show that magnetic compensation in layered ferrimagnetic nanostructures can be effectively tuned by controlled disorder, offering valuable information for the creation of functional magnetic materials with customised thermomagnetic response.
\end{abstract}

\vskip 2cm


\maketitle 
\section{Introduction}
\label{sec_intro}
Functional ferrimagnetic multilayers and magnetic heterostructures have attracted sustained attention due to their tunable thermomagnetic properties and potential applications in spintronic, magnetocaloric, and magnetic storage technologies \cite{Cullity}. In particular, ferrimagnetic systems exhibiting compensation phenomena are of considerable interest because the total magnetization vanishes at a temperature below the magnetic ordering temperature while the constituent magnetic substructures remain magnetically ordered \cite{Neel}. Such compensation behavior emerges from the competition between ferro- and antiferromagnetic intersite interactions associated with distinct magnetic sublattices, sublayers, or subsets of atoms. Owing to their enhanced surface-to-volume ratio and tunable interlayer interactions, \textit{layered} ferrimagnetic systems provide an attractive platform for studying compensation phenomena and engineered magnetic response. Thin ferrimagnets, a subclass of layered ferrimagnetic materials with one dimension significantly smaller than the other two, act as a bridge between monolayer (2D) and bulk (3D) magnetic systems. Recent advances in atomically thin magnetic materials and layered van der Waals heterostructures have further stimulated interest in low-dimensional ferrimagnetic systems with tunable interlayer interactions and controllable magnetic phases \cite{Huang,Burch,Gibertini}.

In recent years, considerable theoretical attention has been devoted to the \textit{compensation effect} in layered ferrimagnetic systems \cite{Diaz1,Diaz2}. The distinct thermal responses of the component magnetic layers may lead to the appearance of a compensation point, i.e., a temperature lower than the critical temperature at which the net magnetization of the system vanishes despite the persistence of long-range magnetic ordering within the individual layers. Layered ferrimagnets with competing ferromagnetic and antiferromagnetic interactions exhibit diverse magnetic responses and complex phase diagrams, making them promising candidates for functional magnetic devices. Such systems have already found applications in giant magnetoresistance (GMR) \cite{Camley}, magneto-optical recording technologies \cite{Connell}, magnetocaloric refrigeration \cite{Phan}, and spintronic architectures \cite{Grunberg}.

Recent experimental progress has enabled the realization of bilayer \cite{Stier}, trilayer \cite{Smits,Leiner}, and multilayered ferrimagnetic systems \cite{Chern,Sankowski,Chung,Samburskaya} with tunable magnetic characteristics. In parallel, advanced thin-film growth techniques such as molecular-beam epitaxy (MBE) \cite{Herman}, metal-organic chemical vapour deposition (MOCVD) \cite{Stringfellow}, pulsed laser deposition (PLD) \cite{SinghRK}, and atomic layer deposition (ALD) \cite{Leskela,George} have played a crucial role in fabricating layered magnetic nanostructures with controlled interfacial and magnetic properties.\\
\indent Alongside experimental developments, theoretical investigations of layered ferrimagnetic systems and magnetic heterostructures have become equally important for understanding and controlling their magnetic properties. Despite substantial progress in analytical and computational techniques, several aspects of low-dimensional magnetism and interlayer exchange interactions remain incompletely understood. A notable example is ${\rm CrI}_{3}$, which exhibits ferromagnetic ordering in the bulk form \cite{Tsubokawa,Dillon}, whereas few-layer stacks display antiferromagnetic interlayer coupling \cite{Huang}. Although electronic structure calculations successfully reproduce the ferromagnetic behaviour of bulk and monolayer ${\rm CrI}_{3}$ \cite{Wang,Sivadas,Lado,Jiang,Zheng}, describing the antiferromagnetic interlayer exchange interaction in bilayer systems remains considerably more challenging \cite{Jang}. Structural reconstruction and stacking-dependent interfacial effects in thin films have been proposed as possible origins of this discrepancy \cite{Thiel,Ubrig}. Furthermore, density functional theory combined with Hubbard corrections [DFT$+U$], within the around-mean-field correction scheme, successfully captures the vertical magnetotransport and experimentally observed magnetoresistance in multilayered ${\rm CrI}_{3}$ heterostructures enclosed between graphene contacts \cite{Sarkar1}. Detailed DFT$+U$ investigations on chromium trihalides ${\rm CrI}_{3}$, ${\rm CrBr}_{3}$, and ${\rm CrCl}_{3}$ have also provided valuable insights into exchange interactions and magnetic ordering temperatures in layered magnetic systems \cite{Sarkar2}. Recent review articles on two-dimensional magnetic materials and van der Waals heterostructures further highlight the growing importance of theoretical modelling in understanding and engineering low-dimensional magnetic phenomena \cite{Burch,Gibertini,Gong}.

Spin-model-based theoretical approaches have proven particularly effective in exploring thermodynamic and magnetic properties of layered ferrimagnetic structures. Mean-field and semi-analytical techniques have been extensively employed to investigate phase transitions, compensation phenomena, and thermodynamic behaviour in multilayer ferrimagnets \cite{Diaz1,Amaral,Franco,Dong,Oliveira1,Amaral2,Franco2,Pelka}. Several analytical approaches, including exactly solvable formulations in specific cases, including transfer-matrix-based methods, decoration-iteration transformations, Bethe-ansatz formulations, and Jordan-Wigner transformations, have also provided important insights into low-dimensional spin systems \cite{Canova,Pereira,Ohanyan,Strecka,Galisova,Ribeiro,Trippe,Zhitomirsky,Topilko}. Technologically relevant properties such as the magnetocaloric effect in layered magnetic systems have also been theoretically investigated using spin models and mean-field approaches \cite{Oliveira2}.

Among numerical techniques, Metropolis Monte Carlo simulation remains one of the most reliable approaches for investigating critical and compensation phenomena in layered magnetic systems with competing interactions and disorder. In earlier studies, the effect of next-nearest-neighbour interactions on compensation behavior and phase transitions in trilayer ferrimagnetic systems was investigated using Monte Carlo simulations on square lattices \cite{Guru1}. Dynamic magnetic response and nonequilibrium compensation phenomena in non-equivalent $ABA$-type trilayer ferrimagnets subjected to oscillating magnetic fields were also explored within the Monte Carlo framework \cite{Guru2}. These studies demonstrate that spin models can provide valuable microscopic insights into compensation phenomena and aid in understanding experimentally relevant magnetic behaviour. Consequently, the ongoing search for magnetically compensated materials with improved functional performance increasingly focuses on the manipulation and control of compensation temperatures \cite{Sandeman,Manosa}, which also forms the central motivation of the present work.\\
\indent \indent Recent developments in diluted magnetic systems and disorder-driven magnetic phenomena in layered ferrimagnets have further highlighted the importance of quenched disorder in tailoring magnetic response. The influence of site dilution on the compensation and critical temperatures of two-dimensional mixed spin systems has been investigated using Monte Carlo simulations in several earlier studies \cite{Aydiner}. In \cite{Diaz3}, the thermodynamic and magnetic properties of a site-diluted spin-$1/2$ Ising multilayer ferrimagnet consisting of non-equivalent planes with dominant intralayer interactions were explored using Monte Carlo simulations combined with the Wolff cluster algorithm. A related Monte Carlo investigation on Ising multilayer ferrimagnets with two distinct non-equivalent planes, one of which is site-diluted, was reported in \cite{Diaz4}. These studies indicate that the introduction of quenched nonmagnetic impurities can substantially modify the magnetic and thermodynamic properties of layered ferrimagnetic systems compared to their pristine counterparts. Consequently, site dilution emerges as a potentially effective route for tuning compensating phases and engineering magnetic response in quasi-two-dimensional ferrimagnetic heterostructures.

Motivated by these developments, in this work we employ Metropolis Monte Carlo simulations to investigate the influence of quenched nonmagnetic disorder on the compensation and critical behaviour of a diluted spin-$1/2$ Ising trilayer ferrimagnet on a square lattice. Particular emphasis is placed on understanding (i) the role of impurity concentration in modifying the compensation and critical temperatures, (ii) the evolution of magnetic phase boundaries and compensation regions with increasing dilution, and (iii) the emergence of phenomenological scaling relations connecting impurity concentration, interaction strengths, and compensation behaviour. By carefully selecting the interaction parameters and introducing controlled nonmagnetic impurities, we demonstrate that controlled site dilution provides an effective route for tailoring the thermomagnetic response and compensation characteristics of layered Ising ferrimagnetic heterostructures.

\indent The rest of the paper is organised as follows. The magnetic model is introduced in Section~\ref{sec_model}. The simulation methodology employed for the trilayer system is presented in Section~\ref{sec_simulation}. The numerical results and discussions are provided in Section~\ref{sec_results}. Finally, the main conclusions of the work are summarised in Section~\ref{sec_summary}.

\section{Theoretical Model}
\label{sec_model}

\begin{figure}[!htb]
		\resizebox{9.0cm}{!}{\includegraphics[angle=0]{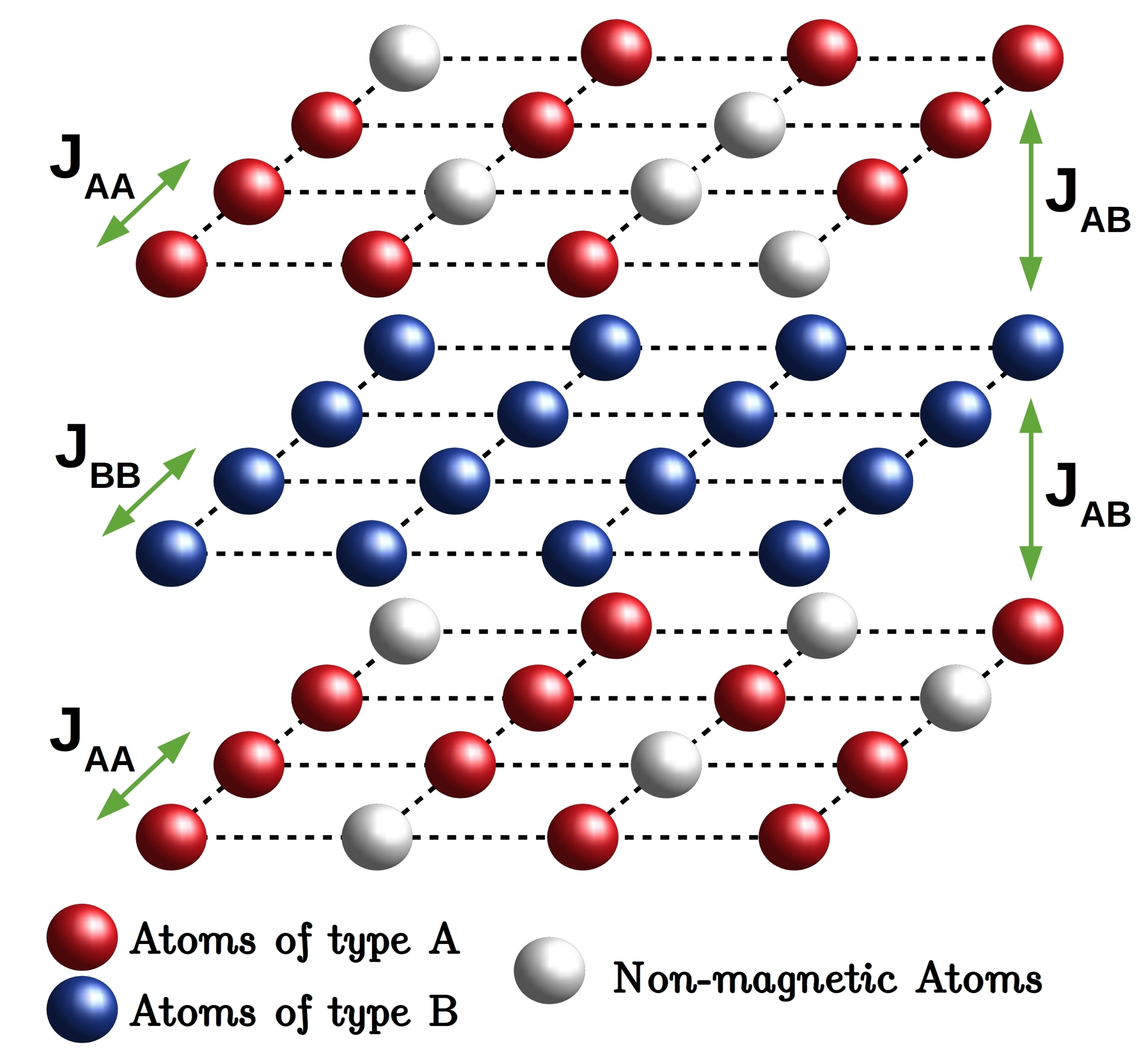}}\\
		{\large \textbf{(a) ABA-type diluted trilayer stacking}}\\\vskip 0.2cm
		\resizebox{9.0cm}{!}{\includegraphics[angle=0]{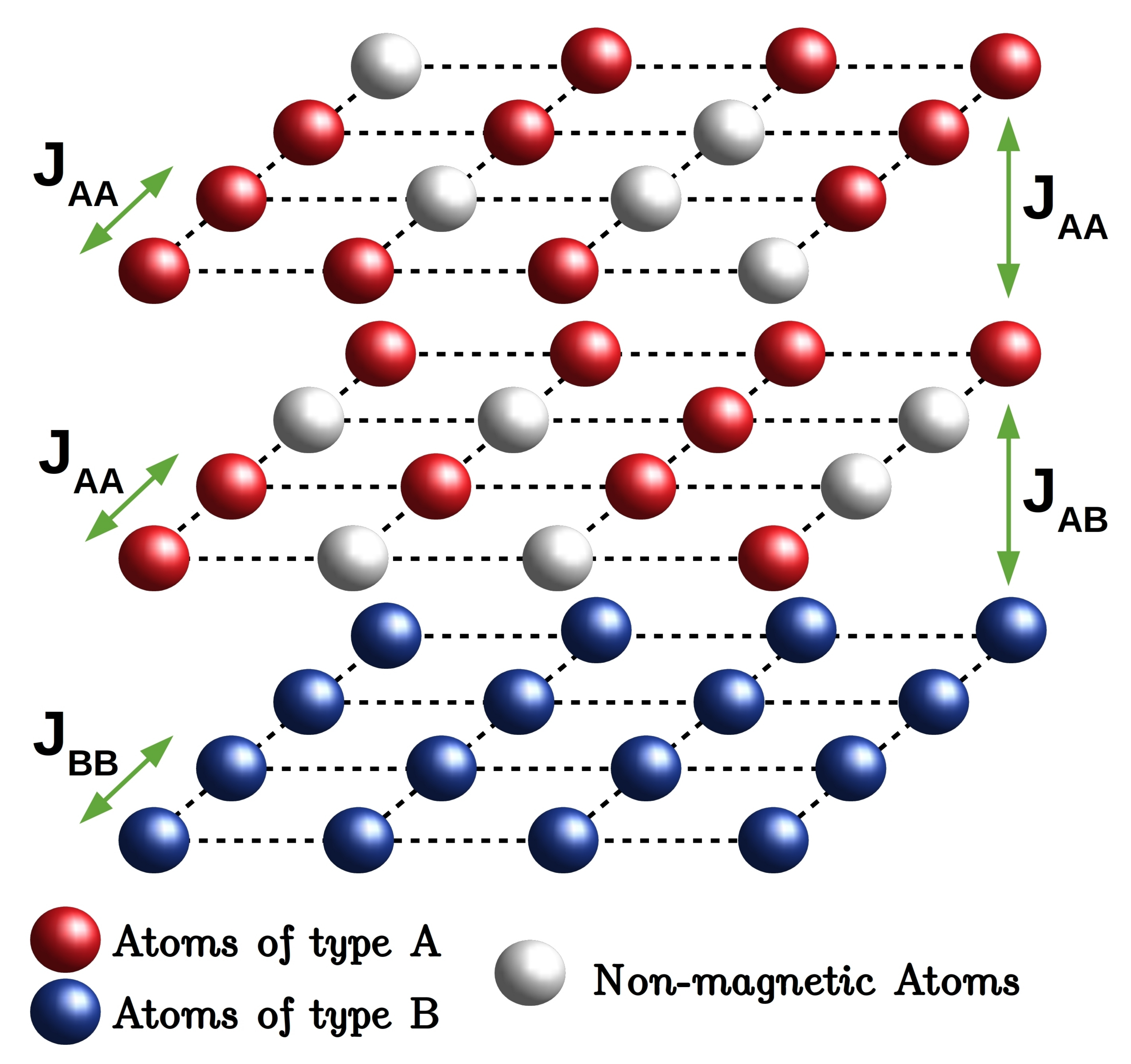}}\\
		{\large \textbf{(b) AAB-type diluted trilayer stacking}}\\
	
	\caption{(Colour online) Schematic representations of the site-diluted trilayer ferrimagnetic systems considered in the present work: (a) $ABA$-type and (b) $AAB$-type layered stackings on square lattices. Nonmagnetic impurities are introduced randomly within the A-layers. The actual simulations are performed on systems of size $N_{\rm sites}=3\times100\times100$.}
\label{fig_1_lattice_structure}
\end{figure}

The present work focuses on site-diluted ferrimagnetic trilayer heterostructures composed of coupled square monolayers occupied by localized spin-$1/2$ magnetic moments. Two distinct stacking configurations are considered, namely the $ABA$-type and $AAB$-type arrangements. In the $ABA$ geometry, the top and bottom layers are formed by A-type magnetic atoms, while the middle layer consists of B-type atoms. In contrast, the $AAB$ structure contains A-type atoms in the top and middle layers, whereas the bottom layer is composed of B-type atoms. In both configurations, quenched nonmagnetic impurities are randomly introduced within the A-layers with concentration $\rho$, while the B-layer remains pristine and possesses the dominant intralayer exchange interaction. Schematic representations of the considered trilayered systems are shown in Fig.~\ref{fig_1_lattice_structure}.

The magnetic interactions are restricted to nearest-neighbour exchange couplings. Ferromagnetic intralayer interactions act between like atoms (A--A and B--B), whereas antiferromagnetic interlayer interactions occur between unlike atoms (A--B). Within the nearest-neighbour Ising formalism \cite{Ising}, the Hamiltonians corresponding to the $ABA$ and $AAB$ trilayer configurations are respectively given by:

\begin{eqnarray}
\nonumber
H_{ABA} &=& - J_{AA} \Bigg[ \sum_{<t,t^{\prime}>} (\xi_{t}S_{t}^{z})(\xi_{t^{\prime}} S_{t^{\prime}}^{z}) \\ 
\nonumber
&+& \sum_{<b,b^{\prime}>} (\xi_{b}S_{b}^{z})(\xi_{b^{\prime}}S_{b^{\prime}}^{z}) \Bigg] - J_{BB} \sum_{<m,m^{\prime}>} S_{m}^{z}S_{m^{\prime}}^{z} \\
&-& J_{AB}\Bigg[ \sum_{<t,m>} (\xi_{t}S_{t}^{z}) S_{m}^{z} +  \sum_{<m,b>} S_{m}^{z} (\xi_{b}S_{b}^{z}) \Bigg]
\label{eq_Hamiltonian_ABA} 
\end{eqnarray}

\begin{eqnarray}
\nonumber
H_{AAB} &=& - J_{AA} \Bigg[ \sum_{<t,t^{\prime}>} (\xi_{t}S_{t}^{z})(\xi_{t^{\prime}} S_{t^{\prime}}^{z}) \\
\nonumber
&+& \sum_{<m,m^{\prime}>} (\xi_{m}S_{m}^{z})(\xi_{m^{\prime}}S_{m^{\prime}}^{z}) \Bigg] - J_{BB} \sum_{<b,b^{\prime}>} S_{b}^{z}S_{b^{\prime}}^{z}  \\
\nonumber
&-& J_{AA} \sum_{<t,m>} (\xi_{t}S_{t}^{z}) (\xi_{m}S_{m}^{z}) \\
&-& J_{AB} \sum_{<m,b>} (\xi_{m}S_{m}^{z}) S_{b}^{z}
\label{eq_Hamiltonian_AAB} 
\end{eqnarray}

Here, $J_{AA}$ and $J_{BB}$ denote the ferromagnetic intralayer exchange interactions within the A- and B-type layers, respectively, while $J_{AB}$ represents the antiferromagnetic interlayer exchange coupling between adjacent A and B layers. Accordingly, the exchange interactions satisfy $J_{AA}>0$, $J_{BB}>0$, and $J_{AB}<0$. The symbols $\langle t,t' \rangle$, $\langle m,m' \rangle$, and $\langle b,b' \rangle$ denote nearest-neighbour summations over the top, middle, and bottom layers, respectively, whereas $\langle t,m \rangle$ and $\langle m,b \rangle$ represent summations over nearest-neighbour pairs belonging to adjacent layers. The governing interaction parameters are the ferromagnetic coupling ratio $J_{AA}/J_{BB}$ and the antiferromagnetic coupling ratio $J_{AB}/J_{BB}$. Throughout this work, the dominant coupling strength $J_{BB}$ is set to unity.

The configurational averages of the site-occupation variables for the diluted A-type layers are given by $\overline{\xi_t}=\overline{\xi_b}=1-\rho$ for the $ABA$ configuration and $\overline{\xi_t}=\overline{\xi_m}=1-\rho$ for the $AAB$ configuration, where $\rho$ denotes the concentration of nonmagnetic impurities. The top and bottom layers do not interact directly across the vertical direction. Periodic boundary conditions are imposed within the layer planes, while open boundary conditions are employed along the stacking direction. The notations adopted here are consistent with those used in Ref.~\cite{Chandra6}.

\section{Methodology}
\label{sec_simulation}
Monte Carlo simulations employing the Metropolis single-spin-flip algorithm \cite{Landau-Binder,Binder-Heermann} are performed to investigate the thermomagnetic behaviour of the site-diluted Ising trilayer systems shown in Fig.~\ref{fig_1_lattice_structure}. The simulations are carried out on lattices of total size $N_{\rm sites}=3\times100\times100$, where each layer consists of a $100\times100$ square lattice. Based on the finite-size considerations discussed in Section~\ref{appendix_comp}, the chosen system size is found to provide statistically reliable estimates of the relevant thermodynamic quantities and compensation behaviour.

The localized spin variables correspond to Ising spin-$1/2$ moments with $S_{iz}=\pm1/2$. For each disorder realization, quenched nonmagnetic impurities are randomly distributed over the A-type layers with concentration $\rho$, while the B-layer remains magnetically pristine. The simulations are initiated from random spin configurations corresponding to a high-temperature disordered state, where equal populations of $S_{iz}=+1/2$ and $S_{iz}=-1/2$ are assigned to the occupied magnetic sites. During the simulation, a trial configuration is generated by randomly selecting and attempting to flip a single spin. The trial move is accepted according to the standard Metropolis transition probability

\begin{equation}
p=\min \left\{1,e^{-\beta \Delta E}\right\},
\end{equation}

where $\Delta E$ denotes the energy difference between the trial and the current spin configurations, and $\beta=(k_{B}T)^{-1}$. One Monte Carlo sweep (MCS) consists of $3L^{2}$ attempted single-spin updates and serves as the fundamental unit of simulation time throughout this work.

At each temperature step, the first $2\times10^{4}$ MCS are discarded for equilibration. Statistical measurements are subsequently performed over the next $8\times10^{4}$ MCS. To reduce autocorrelation effects, physical observables are sampled after every $800$ MCS, yielding $100$ effectively uncorrelated configurations for statistical averaging. The final spin configuration obtained at a given temperature is used as the initial configuration for the subsequent lower temperature during the cooling procedure.

The dimensionless temperature is expressed in units of $J_{BB}/k_{B}$, where the dominant ferromagnetic exchange interaction of the B-layer, $J_{BB}$, sets the energy scale and is fixed to unity throughout the simulations. The remaining interaction strengths, namely $J_{AA}$ and $J_{AB}$, are measured relative to $J_{BB}$. Periodic boundary conditions are imposed within the layer planes (along the $x$ and $y$ directions), whereas open boundary conditions are employed along the stacking direction ($z$ axis). Seven distinct values of the interaction ratios $J_{AA}/J_{BB}$ and $\left|J_{AB}/J_{BB}\right|$, ranging from $0.04$ to $1.0$ in steps of $0.16$, are considered in the present study. We now define the thermodynamic observables used to characterize the magnetic behaviour of the diluted Ising trilayer systems.

\textbf{(1) Sublattice magnetizations:}

The instantaneous sublattice magnetization corresponding to the $q$-th layer ($q=t,m,b$) at temperature $T$ is defined as

\begin{equation}
m_{qi}(T)=\dfrac{1}{L^{2}}\sum_{x,y=1}^{L}\left(S_{qi}^{z}(T)\right)_{xy},
\end{equation}

where $i$ labels the sampled Monte Carlo configuration. The thermal average of the sublattice magnetization is obtained from $N$ statistically uncorrelated configurations according to

\begin{equation}
\langle m_{q}(T)\rangle=\dfrac{1}{N}\sum_{i=1}^{N}m_{qi}(T).
\end{equation}

The total magnetization per site of the trilayer system, serving as the order parameter of the ferrimagnet, is defined as

\begin{equation}
m_{\rm tot}(T)=\dfrac{1}{3}\left[\langle m_{t}(T)\rangle+\langle m_{m}(T)\rangle+\langle m_{b}(T)\rangle\right].
\end{equation}

\textbf{(2) Internal energy per site:}

The thermal average of the internal energy per site is calculated as

\begin{equation}
\varepsilon(T)=\dfrac{\langle H(T)\rangle}{3L^{2}},
\end{equation}

where $H(T)$ is determined using Eqs.~(\ref{eq_Hamiltonian_ABA}) and (\ref{eq_Hamiltonian_AAB}).

\textbf{(3) Magnetic susceptibility and specific heat:}

The magnetic susceptibility per site is evaluated from the fluctuation of the total magnetization according to

\begin{equation}
\chi(T)=3L^{2}\dfrac{\left[\Delta m_{\rm tot}(T)\right]^{2}}{k_{B}T},
\end{equation}

where $\Delta m_{\rm tot}(T)$ denotes the fluctuation of the total magnetization. Similarly, the specific heat per site is determined from the fluctuation of the internal energy as

\begin{equation}
C(T)=3L^{2}\dfrac{\left[\Delta \varepsilon(T)\right]^{2}}{k_{B}T^{2}}.
\end{equation}

For a generic observable $A$, the associated fluctuation is defined by

\begin{equation}
\Delta A(T)=\sqrt{\dfrac{1}{N}\sum_{i=1}^{N}\left[A_{i}(T)-\langle A(T)\rangle\right]^{2}},
\end{equation}

where $A_{i}(T)$ denotes the value of the observable measured for the $i$-th sampled configuration.

\textbf{(4) Fourth-order Binder cumulant:}

To characterize the critical behaviour of the system, the fourth-order Binder cumulant of the total magnetization is calculated as \cite{Binder-Heermann}

\begin{equation}
\left[U_{4}(T)\right]_{L}
=
1-\dfrac{\langle m_{\rm tot}^{4}(T)\rangle_{L}}
{3\langle m_{\rm tot}^{2}(T)\rangle_{L}^{2}}.
\end{equation}

A lattice size of $L=100$ is employed throughout the present study. For each temperature and parameter set, thermal averages are additionally performed over $10$ independent disorder realizations corresponding to different microscopic impurity distributions with identical macroscopic concentrations. Statistical uncertainties are estimated using the Jackknife resampling technique \cite{Newman}. Except in the immediate vicinity of the critical region, the resulting error bars are generally smaller than the symbol sizes used in the figures.

\section{Results}
\label{sec_results}

\subsection{Thermodynamic response}

We first discuss the thermomagnetic response of the diluted Ising trilayer systems for three representative impurity concentrations, namely $\rho=\{0.00,0.20,0.40\}$. For both the $ABA$ and $AAB$ stackings, the magnetization profiles are found to depend sensitively on three principal control parameters: (i) the ferromagnetic coupling ratio (FR), $J_{AA}/J_{BB}$, (ii) the antiferromagnetic coupling ratio (AFR), $J_{AB}/J_{BB}$, and (iii) the concentration of quenched nonmagnetic impurities, $\rho$. In the present subsection, particular emphasis is placed on the influence of site dilution on the evolution of the magnetization profiles classified according to the Néel nomenclature \cite{Strecka}.

\begin{figure*}[!htb]
	\begin{center}
		\begin{tabular}{c}
			
		\resizebox{9.0cm}{!}{\includegraphics[angle=0]{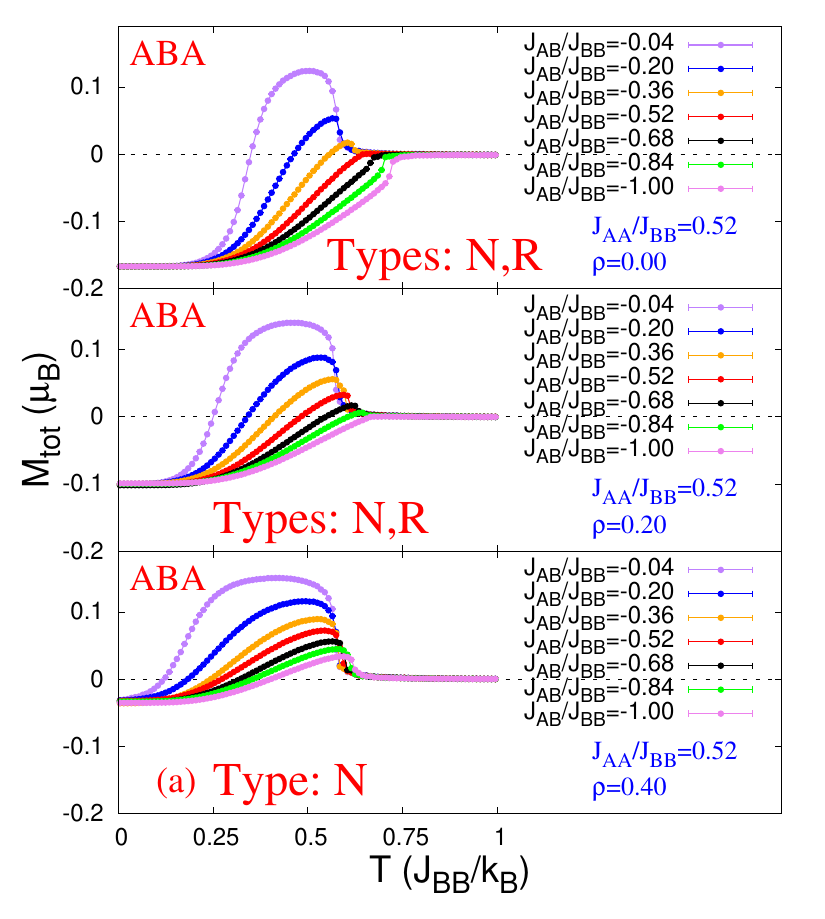}}
		\resizebox{9.0cm}{!}{\includegraphics[angle=0]{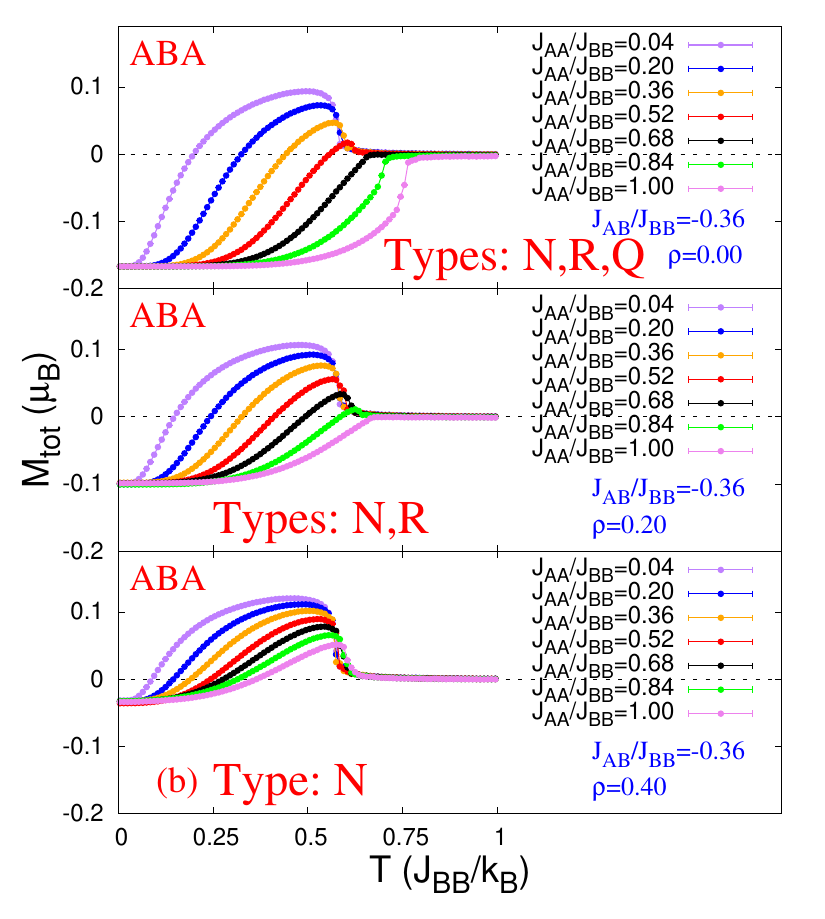}}\\
			
		\resizebox{9.0cm}{!}{\includegraphics[angle=0]{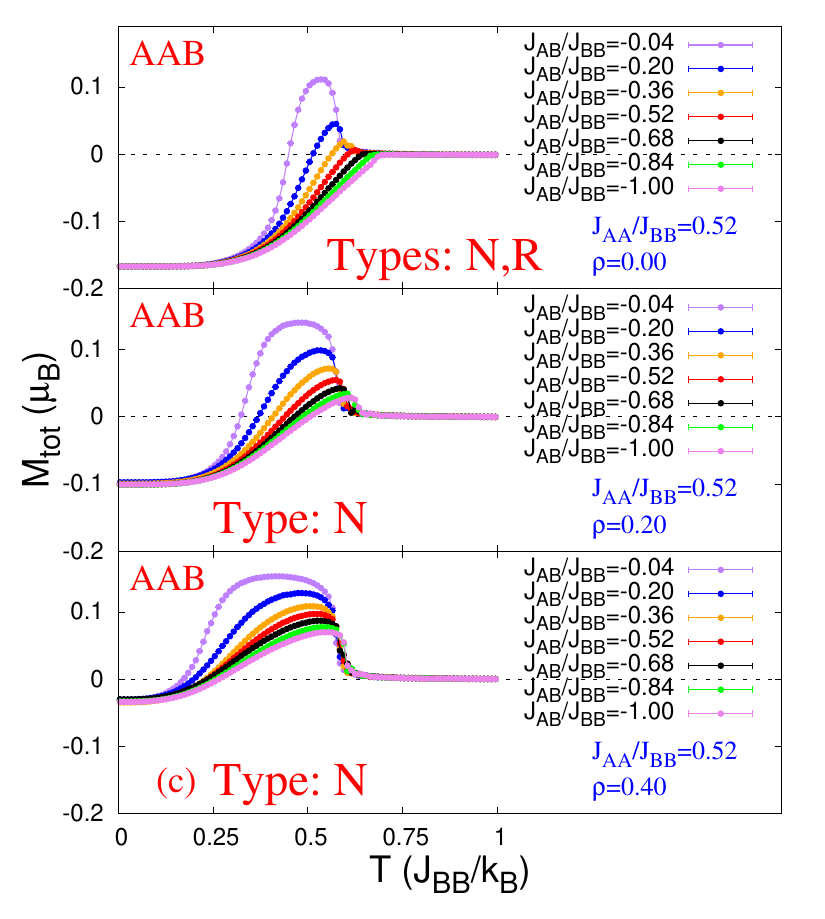}}
		\resizebox{9.0cm}{!}{\includegraphics[angle=0]{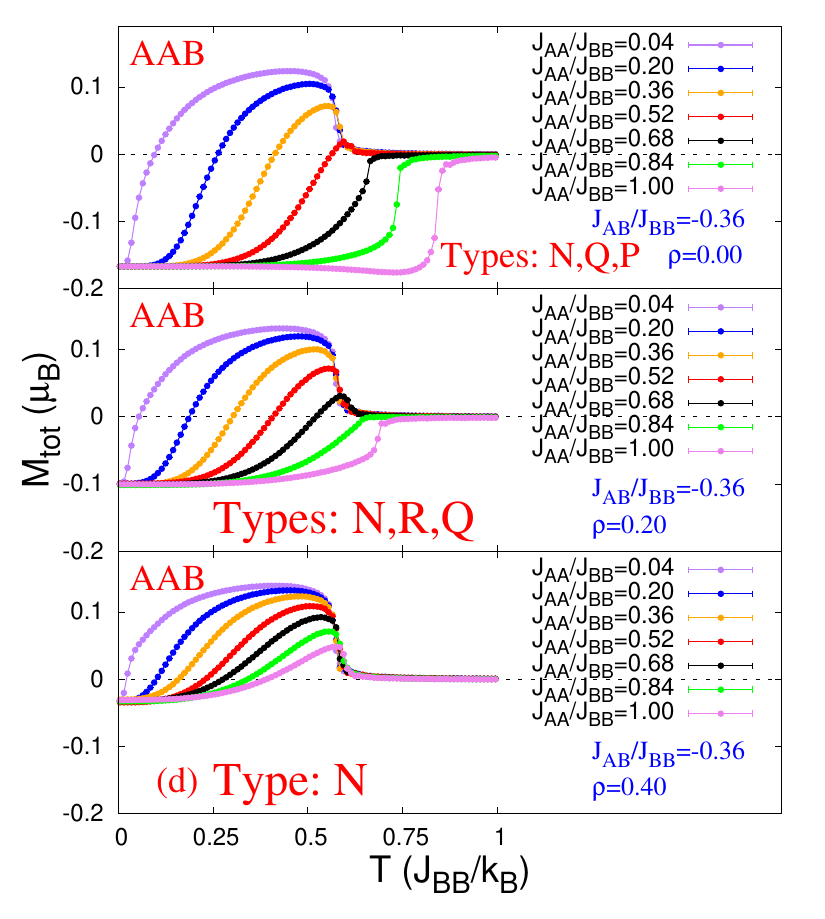}}
			
		\end{tabular}
		\caption{(Colour online) Temperature dependence of the total magnetization, $m_{\rm tot}$, for different concentrations of nonmagnetic impurities, $\rho$, in diluted Ising trilayer systems. Panels (a) and (b) correspond to the $ABA$ stacking for fixed $J_{AA}/J_{BB}=0.52$ and fixed $J_{AB}/J_{BB}=-0.36$, respectively. Panels (c) and (d) show the corresponding results for the $AAB$ stacking.}
		\label{fig_2_mag_temp_rho}
	\end{center}
\end{figure*}

For the $ABA$ stacking, Fig.~\ref{fig_2_mag_temp_rho}(a) shows the temperature dependence of the total magnetization for fixed ferromagnetic coupling ratio $J_{AA}/J_{BB}=0.52$ and varying antiferromagnetic coupling strengths. In the pristine system ($\rho=0.00$), both compensating and non-compensating magnetization profiles are observed. With increasing impurity concentration, the number of N-type magnetization curves progressively increases, and for $\rho=0.40$ all the accessible magnetization profiles exhibit compensation behaviour. A similar tendency is observed in Fig.~\ref{fig_2_mag_temp_rho}(b), where the antiferromagnetic ratio is fixed at $J_{AB}/J_{BB}=-0.36$ while the ferromagnetic interaction strength is varied. In this case, pristine systems display N-, R-, and Q-type magnetization profiles, whereas increasing site dilution suppresses the non-compensating phases, leading predominantly to N-type behaviour at higher impurity concentrations.

A qualitatively similar evolution is obtained for the $AAB$ trilayer geometry. In Fig.~\ref{fig_2_mag_temp_rho}(c), corresponding to fixed $J_{AA}/J_{BB}=0.52$, both N- and R-type magnetization profiles are present in the pristine system. However, increasing dilution strongly favours the emergence of compensating N-type behaviour, and for $\rho=0.20$ and $\rho=0.40$ all accessible magnetization curves exhibit compensation points. Likewise, in Fig.~\ref{fig_2_mag_temp_rho}(d), obtained for fixed $J_{AB}/J_{BB}=-0.36$, pristine systems display N-, Q-, and P-type profiles. Increasing impurity concentration gradually transforms the magnetic response through intermediate R-type behaviour and ultimately stabilizes N-type compensation profiles at sufficiently large dilution. In all investigated cases, both the compensation temperature and the magnetic transition temperature shift toward higher values with increasing ferromagnetic and antiferromagnetic exchange strengths when the impurity concentration is kept fixed. These results demonstrate that controlled site dilution provides an additional mechanism for tuning compensation behaviour and modifying the thermomagnetic response of Ising trilayer ferrimagnets.

\begin{figure*}[!htb]
	\begin{center}
		\begin{tabular}{c}
			
			\resizebox{9.0cm}{!}{\includegraphics[angle=0]{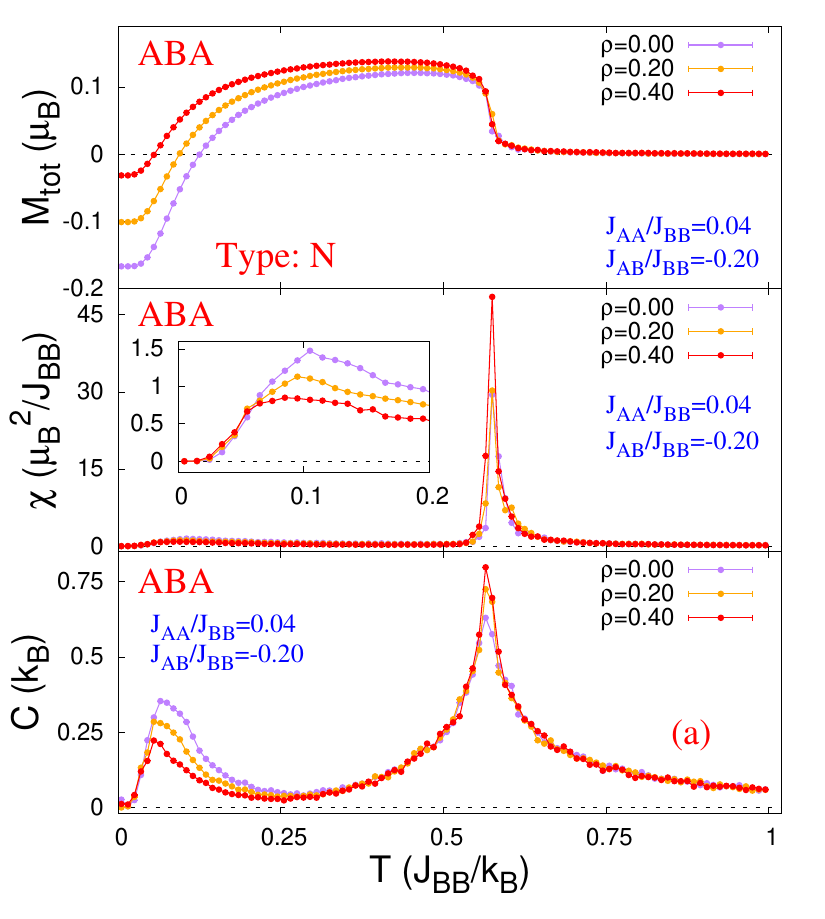}}
			\resizebox{9.0cm}{!}{\includegraphics[angle=0]{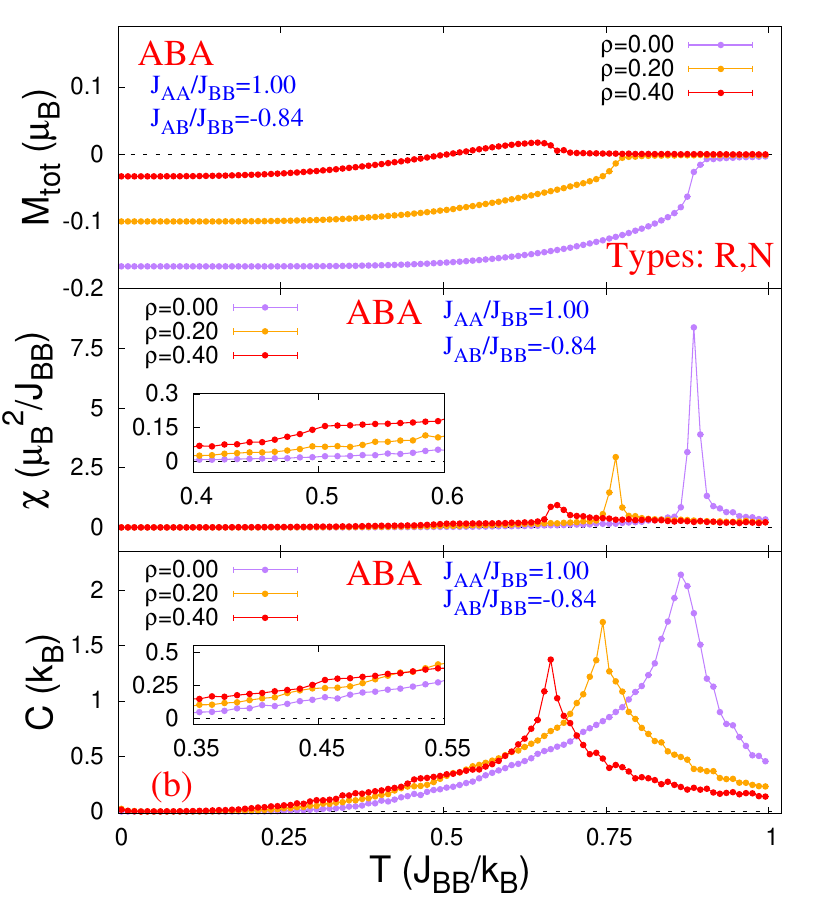}}\\
			
			\resizebox{9.0cm}{!}{\includegraphics[angle=0]{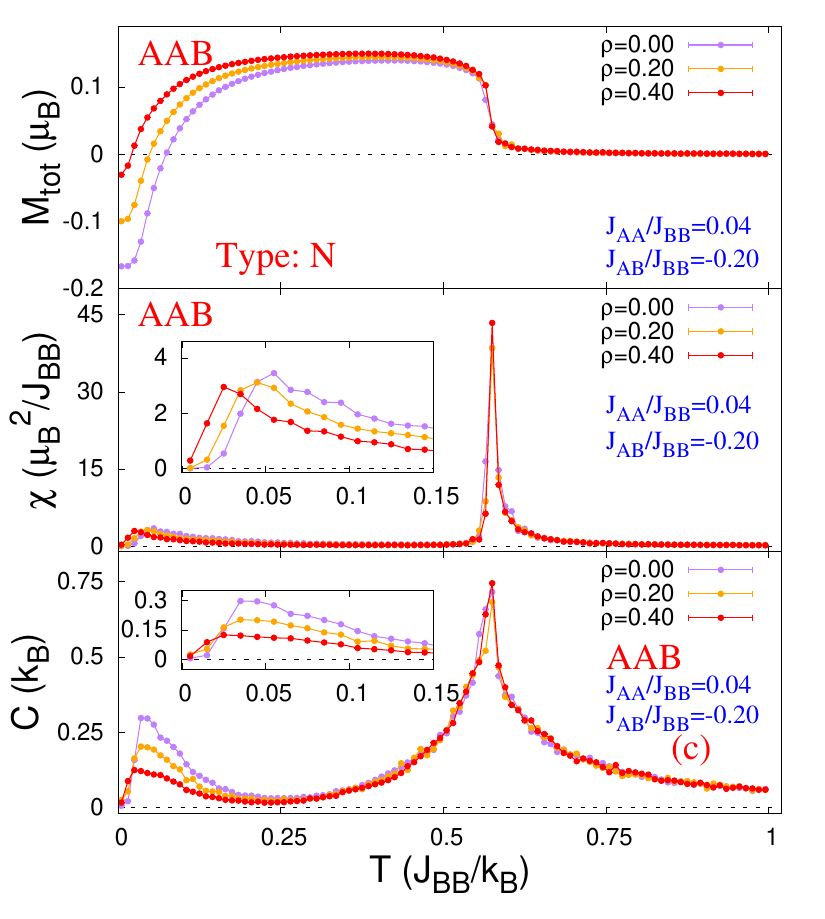}}
			\resizebox{9.0cm}{!}{\includegraphics[angle=0]{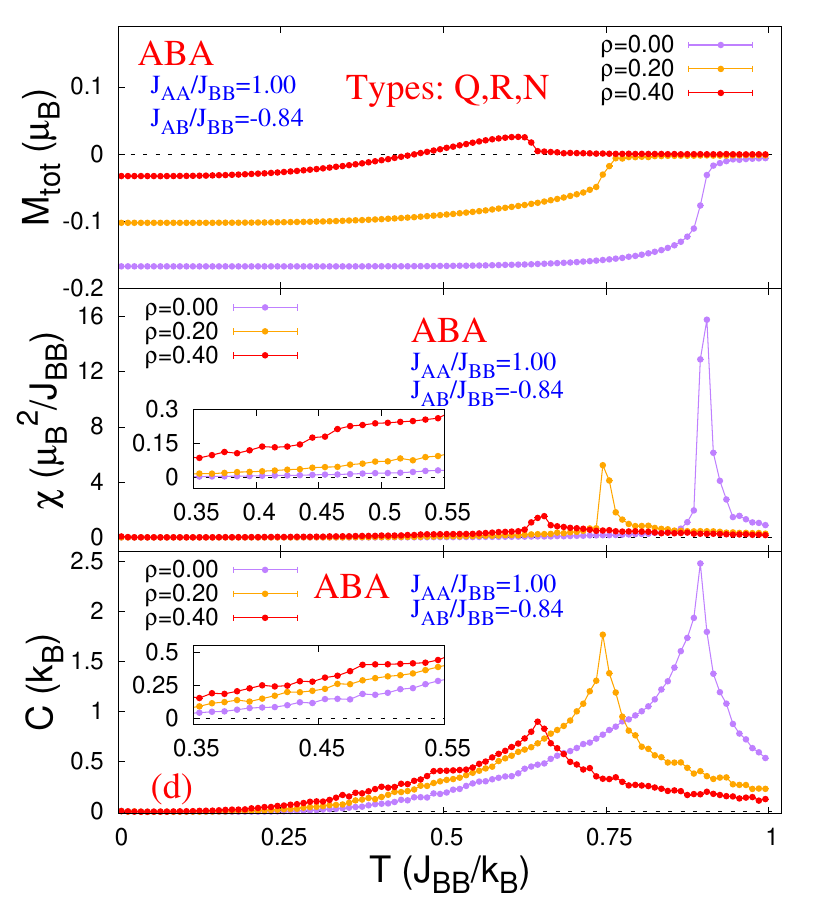}}
			
		\end{tabular}
		\caption{(Colour online) Temperature dependence of the total magnetization $m_{\rm tot}$, magnetic susceptibility $\chi$, and specific heat $C$ for different impurity concentrations $\rho$ in diluted Ising trilayer ferrimagnets. Panels (a) and (b) correspond to the $ABA$ stacking for $(J_{AA}/J_{BB},J_{AB}/J_{BB})=(0.04,-0.20)$ and $(1.00,-0.84)$, respectively. Panels (c) and (d) show the corresponding results for the $AAB$ stacking. Insets highlight the thermodynamic response near the compensation region.}
		\label{fig_3_mag_chi_sph_rho}
	\end{center}
\end{figure*}

To further characterize the influence of quenched disorder on the thermodynamic response, Fig.~\ref{fig_3_mag_chi_sph_rho} presents the evolution of the magnetization profiles, magnetic susceptibility, and specific heat for selected combinations of ferromagnetic and antiferromagnetic coupling strengths as the impurity concentration is varied. For the $ABA$ trilayer configuration, Fig.~\ref{fig_3_mag_chi_sph_rho}(a), corresponding to $J_{AA}/J_{BB}=0.04$ and $J_{AB}/J_{BB}=-0.20$, exhibits exclusively N-type magnetization behaviour for all impurity concentrations considered. The compensation temperature systematically shifts toward lower temperatures with increasing dilution. The corresponding susceptibility and specific heat curves display pronounced peaks in the vicinity of the magnetic transition temperature, signalling critical behaviour, while broadened secondary features appear near the compensation temperature, as highlighted in the inset. The persistence of N-type behaviour over the entire dilution range indicates that weak ferromagnetic intralayer coupling combined with moderate antiferromagnetic interaction favours robust compensation characteristics even in the presence of substantial quenched disorder.

A more intricate evolution is observed in Fig.~\ref{fig_3_mag_chi_sph_rho}(b), obtained for $J_{AA}/J_{BB}=1.00$ and $J_{AB}/J_{BB}=-0.84$. In this case, the pristine system exhibits an R-type magnetization profile without compensation. Increasing the impurity concentration induces a crossover toward compensating N-type behaviour, which becomes fully established for $\rho=0.40$. Correspondingly, the susceptibility and specific heat exhibit singular peaks near the transition temperature for all concentrations, whereas additional broadened structures associated with the compensation point emerge only in the N-type regime. This behaviour suggests that site dilution weakens the effective ferromagnetic contribution of the diluted A-layers sufficiently to stabilize compensation behaviour in parameter regions where it is absent in the pristine system.

For the $AAB$ trilayer geometry, Fig.~\ref{fig_3_mag_chi_sph_rho}(c), corresponding to $J_{AA}/J_{BB}=0.04$ and $J_{AB}/J_{BB}=-0.20$, also displays N-type magnetization profiles throughout the investigated dilution range. Increasing impurity concentration shifts the compensation temperature toward lower values while preserving the overall thermodynamic characteristics. Pronounced peaks in the susceptibility and specific heat continue to identify the transition temperature, accompanied by weaker broadened anomalies near the compensation point. Compared with the corresponding $ABA$ configuration, the compensation behaviour in the $AAB$ stacking remains relatively stable against disorder due to the asymmetric arrangement of the magnetic layers.

Finally, Fig.~\ref{fig_3_mag_chi_sph_rho}(d), corresponding to $J_{AA}/J_{BB}=1.00$ and $J_{AB}/J_{BB}=-0.84$, reveals a sequence of disorder-driven transformations in the magnetization profiles. The pristine system initially exhibits a Q-type response without compensation, which evolves into an R-type profile at intermediate dilution and finally into an N-type compensating phase for $\rho=0.40$. The associated thermodynamic response functions consistently exhibit sharp maxima near the magnetic transition temperature, while broadened features near the compensation point appear only when compensation behaviour is present. These observations collectively demonstrate that quenched site dilution can strongly modify the thermodynamic response and compensation characteristics of diluted Ising trilayer ferrimagnets.

\begin{figure*}[!htb]
	\begin{center}
		\begin{tabular}{c}
			
			\resizebox{9.0cm}{!}{\includegraphics[angle=0]{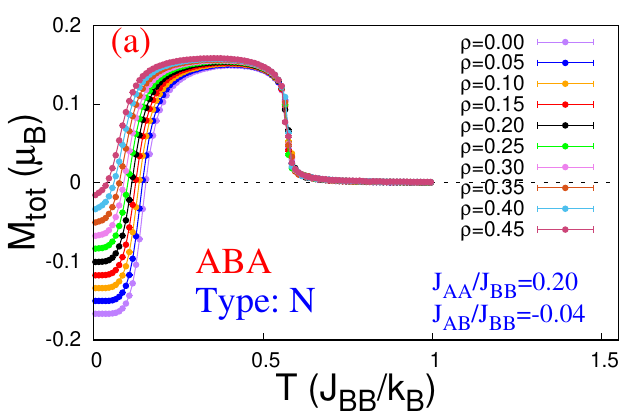}}
			\resizebox{9.0cm}{!}{\includegraphics[angle=0]{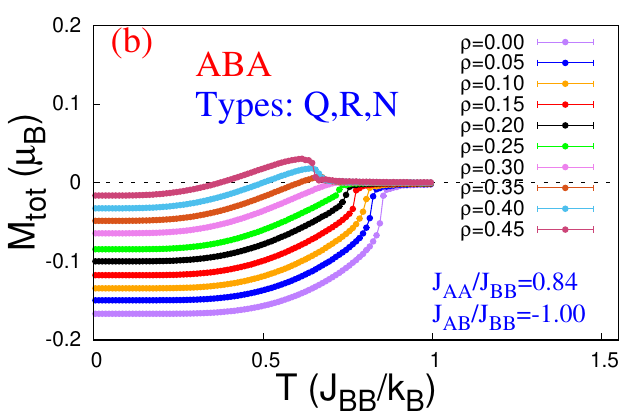}}\\
			
			\resizebox{9.0cm}{!}{\includegraphics[angle=0]{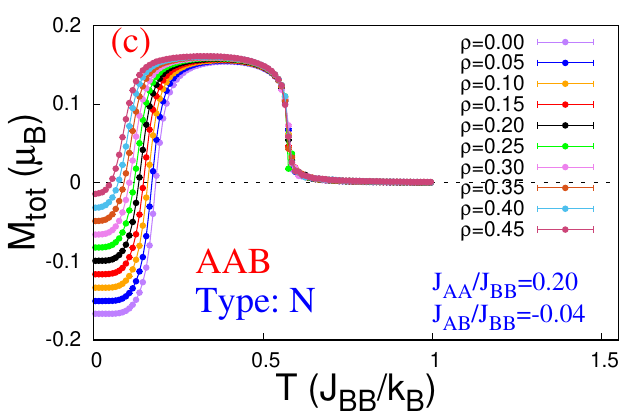}}
			\resizebox{9.0cm}{!}{\includegraphics[angle=0]{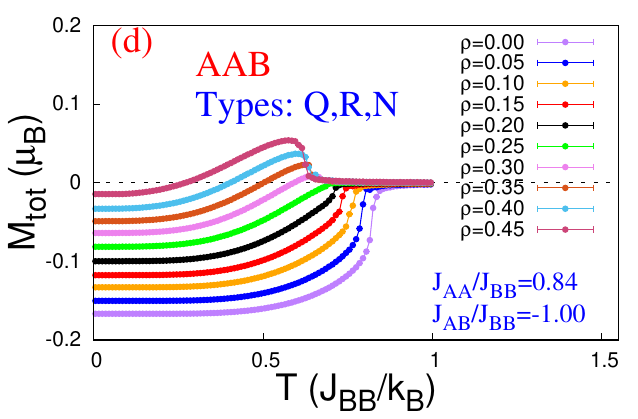}}
			
		\end{tabular}
		\caption{(Colour online) Temperature dependence of the total magnetization $m_{\rm tot}$ for different concentrations of quenched nonmagnetic impurities $\rho$ in diluted Ising trilayer ferrimagnets. Panels (a) and (b) correspond to the $ABA$ stacking for $(J_{AA}/J_{BB},J_{AB}/J_{BB})=(0.20,-0.04)$ and $(0.84,-1.00)$, respectively. Panels (c) and (d) show the corresponding results for the $AAB$ stacking.}
		\label{fig_4_mag_rhomag}
	\end{center}
\end{figure*}

At this stage, it is instructive to examine several representative cases in greater detail in order to understand how the magnetization profiles evolve continuously with increasing concentration of quenched nonmagnetic impurities and how disorder-induced compensation behaviour emerges in the diluted Ising trilayer systems. For the $ABA$ trilayer geometry, Fig.~\ref{fig_4_mag_rhomag}(a) corresponds to relatively weak exchange interactions with $J_{AA}/J_{BB}=0.20$ and $J_{AB}/J_{BB}=-0.04$. In this regime, all magnetization profiles belong to the N-type class and exhibit well-defined compensation temperatures. Increasing the impurity concentration systematically shifts the compensation point toward lower temperatures while preserving the qualitative nature of the magnetic response. The gradual suppression of the compensation temperature reflects the weakening of the effective ferromagnetic contribution arising from the diluted A-layers.

Different behaviour is observed in Fig.~\ref{fig_4_mag_rhomag}(b), corresponding to stronger exchange couplings with $J_{AA}/J_{BB}=0.84$ and $J_{AB}/J_{BB}=-1.00$. In the pristine and weakly diluted regimes, the system predominantly exhibits Q- and R-type magnetization profiles without compensation. However, increasing the concentration of nonmagnetic impurities progressively transforms these non-compensating phases into compensating N-type profiles. For sufficiently large dilution ($\rho \gtrsim 0.35$), stable compensation points emerge in the magnetization curves. This constitutes a particularly important observation of the present work, demonstrating that compensation behaviour can be induced solely through controlled quenched disorder without modifying the exchange interaction strengths.

A similar disorder-driven evolution is observed for the $AAB$ trilayer configuration. In Fig.~\ref{fig_4_mag_rhomag}(c), corresponding again to weak exchange interactions with $J_{AA}/J_{BB}=0.20$ and $J_{AB}/J_{BB}=-0.04$, all magnetization curves remain of N-type and display compensation temperatures shifting toward lower values with increasing impurity concentration. The overall thermomagnetic response closely resembles that observed for the corresponding $ABA$ geometry, although quantitative differences in the compensation temperatures arise from the distinct layer arrangements.

More strikingly, Fig.~\ref{fig_4_mag_rhomag}(d), corresponding to strong exchange interactions with $J_{AA}/J_{BB}=0.84$ and $J_{AB}/J_{BB}=-1.00$, reveals that increasing site dilution continuously drives the system from non-compensating Q- and R-type phases toward compensating N-type behaviour. Beyond a threshold impurity concentration, stable compensation points emerge even though the pristine system does not exhibit compensation. These results clearly establish that quenched site dilution provides an effective mechanism for engineering compensation behaviour in both $ABA$ and $AAB$ Ising trilayer ferrimagnets.

\begin{figure*}[!htb]
	\begin{center}
		\begin{tabular}{c}
			
			\resizebox{9.0cm}{!}{\includegraphics[angle=0]{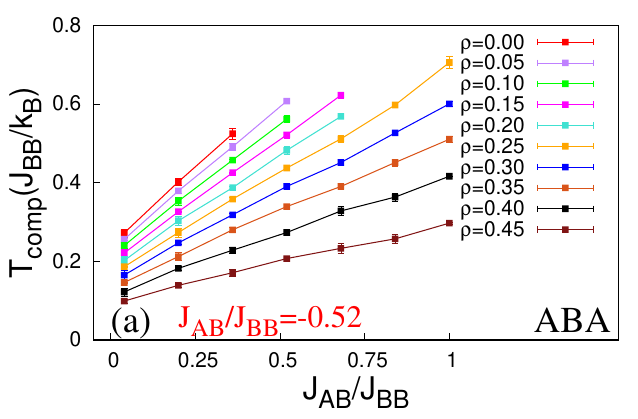}}
			\resizebox{9.0cm}{!}{\includegraphics[angle=0]{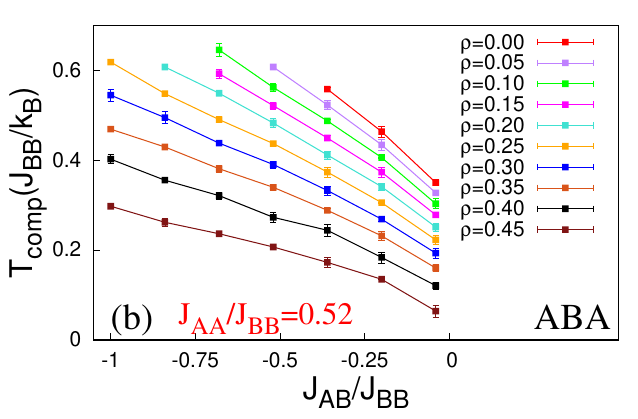}}\\
			
			\resizebox{9.0cm}{!}{\includegraphics[angle=0]{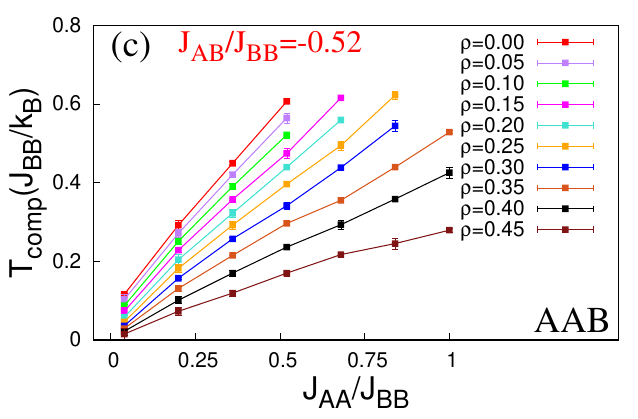}}
			\resizebox{9.0cm}{!}{\includegraphics[angle=0]{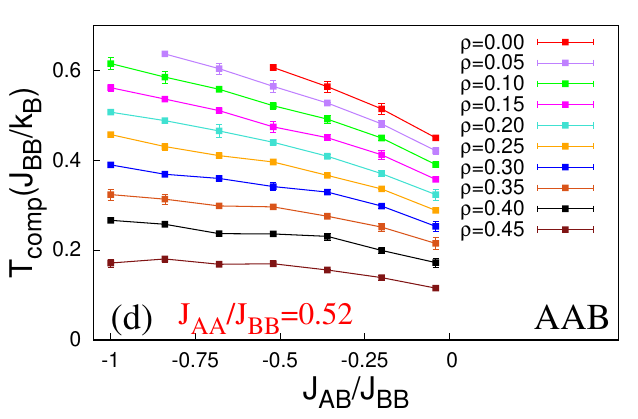}}
			
		\end{tabular}
		\caption{(Colour online) Compensation temperature $T_{\rm comp}$ as a function of exchange interaction ratios for different impurity concentrations $\rho$ in diluted Ising trilayer ferrimagnets. Panels (a) and (b) correspond to the $ABA$ stacking, showing $T_{\rm comp}$ versus $J_{AA}/J_{BB}$ and $J_{AB}/J_{BB}$, respectively. Panels (c) and (d) show the corresponding results for the $AAB$ stacking.}
		\label{fig_5_tcomp_cr_rho}
	\end{center}
\end{figure*}

We are now in a position to analyse how the emergence of compensation behaviour depends systematically on the concentration of quenched nonmagnetic impurities. Figure~\ref{fig_5_tcomp_cr_rho} summarizes the evolution of the compensation temperature in the interaction-parameter space for different impurity concentrations. For example, in Fig.~\ref{fig_5_tcomp_cr_rho}(a), corresponding to the $ABA$ trilayer with fixed antiferromagnetic coupling ratio $J_{AB}/J_{BB}=-0.52$, the pristine system ($\rho=0.00$) exhibits compensation only within a restricted range of ferromagnetic interaction strengths. Upon introducing a small concentration of nonmagnetic impurities, compensation behaviour begins to appear over a wider interval of $J_{AA}/J_{BB}$. As the impurity concentration increases further, the compensation region expands progressively toward stronger interaction strengths. For sufficiently large dilution ($\rho\gtrsim0.25$), compensation behaviour becomes accessible throughout nearly the entire investigated parameter range. An analogous evolution is observed in the remaining panels of Fig.~\ref{fig_5_tcomp_cr_rho}. Increasing the concentration of quenched disorder systematically modifies the compensation boundaries in the Hamiltonian parameter space for both the $ABA$ and $AAB$ trilayer geometries. The impurity concentration therefore acts as an additional tuning parameter controlling the onset and stability of compensation behaviour. These results demonstrate that the phase boundaries associated with compensating and non-compensating magnetic phases are strongly disorder dependent and can be continuously engineered through controlled site dilution.

\subsection{Phase diagrams}

The preceding analysis has already demonstrated that quenched nonmagnetic impurities can fundamentally modify the compensation behaviour of diluted Ising trilayer ferrimagnets. In particular, the introduction of disorder generates compensation points for several interaction-parameter combinations that remain non-compensating in the pristine systems. It therefore becomes important to investigate systematically how the compensation and non-compensation regions evolve within the Hamiltonian parameter space.

\begin{figure*}[!htb]
	\begin{center}
		\begin{tabular}{c}
			
		\resizebox{8.5cm}{!}{\includegraphics[angle=0]{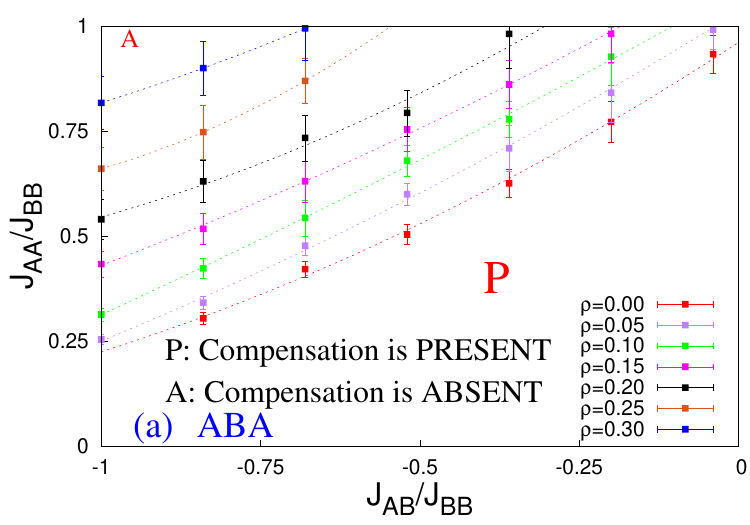}}
		\resizebox{8.5cm}{!}{\includegraphics[angle=0]{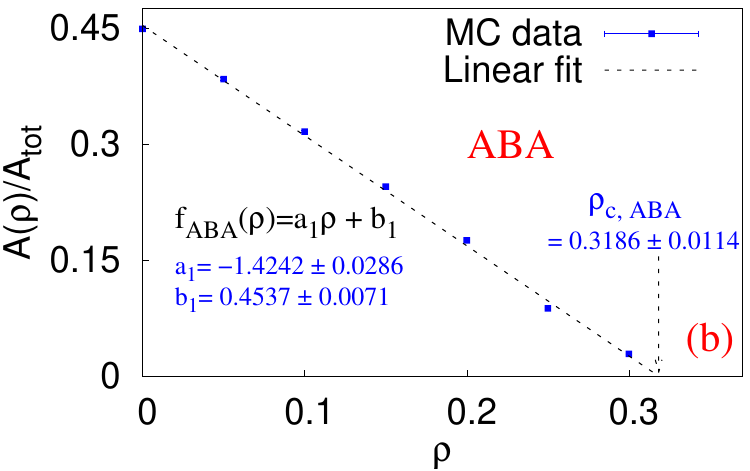}}\\
			
		\resizebox{8.5cm}{!}{\includegraphics[angle=0]{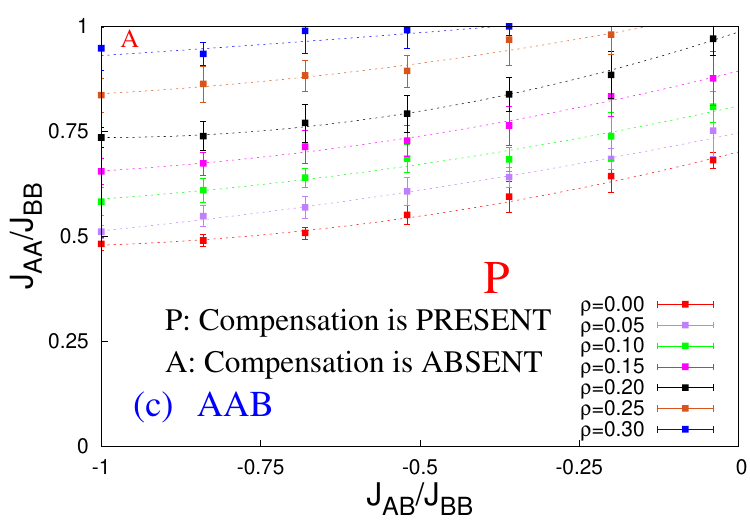}}
		\resizebox{8.5cm}{!}{\includegraphics[angle=0]{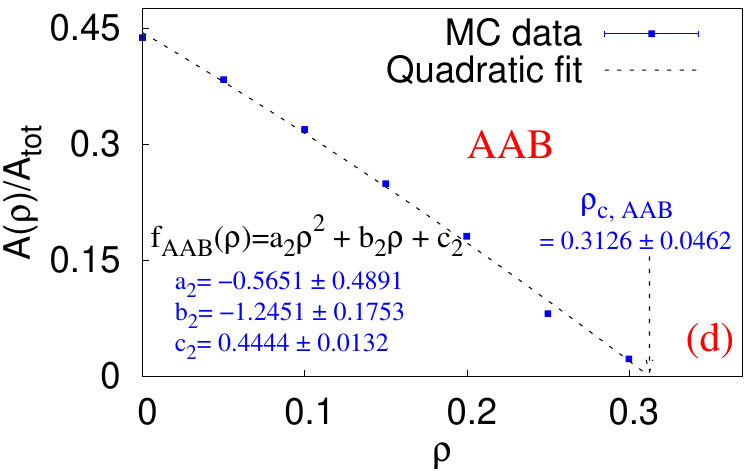}}
			
		\end{tabular}
	\caption{(Colour online) Phase diagrams in the $(J_{AB}/J_{BB}\times J_{AA}/J_{BB})$ interaction plane for diluted Ising trilayer ferrimagnets with different concentrations of quenched nonmagnetic impurities $\rho$. Panels (a) and (c) correspond to the $ABA$ and $AAB$ trilayer configurations, respectively. The corresponding fractional areas of the non-compensating phase region, $A(\rho)/A_{\mathrm{tot}}$, are shown in panels (b) and (d).}
	\label{fig_6_phase_diagram_fit}
	\end{center}
\end{figure*}

Following the procedure discussed in Ref.~\cite{Chandra2}, we construct the phase diagrams in the $(J_{AB}/J_{BB}\times J_{AA}/J_{BB})$ interaction plane for different concentrations of quenched nonmagnetic impurities. The resulting phase boundaries for the $ABA$ and $AAB$ trilayer geometries are presented in Fig.~\ref{fig_6_phase_diagram_fit}(a) and Fig.~\ref{fig_6_phase_diagram_fit}(c), respectively. In these diagrams, the phase curves divide the interaction plane into two distinct thermomagnetic regions. The regions labelled by $A$ correspond to parameter combinations for which no compensation temperature exists, whereas the regions labelled by $P$ represent interaction strengths that yield compensating magnetization behaviour.

A clear disorder-induced evolution of the phase boundaries is observed for both trilayer configurations. As the impurity concentration increases, the non-compensating region progressively shrinks, while the compensating region expands toward larger portions of the interaction plane. Beyond a sufficiently large impurity concentration, nearly all investigated combinations of coupling strengths exhibit compensation behaviour. These results demonstrate that quenched site dilution acts as an efficient tuning mechanism for controlling the compensation characteristics of layered ferrimagnetic systems.

To quantify this evolution, we calculate the fractional area of the non-compensating region, denoted by $A(\rho)/A_{\mathrm{tot}}$, as a function of impurity concentration $\rho$. Here, $A(\rho)$ represents the area of the non-compensating phase region within the investigated interaction domain, while $A_{\mathrm{tot}}$ denotes the total explored parameter-space area. Required areas are estimated numerically using Monte Carlo integration techniques \cite{Krauth}. Resulting impurity-dependent behaviour of the fractional non-compensating area is presented in Fig.~\ref{fig_6_phase_diagram_fit}(b) for the $ABA$ and in Fig.~\ref{fig_6_phase_diagram_fit}(d) for the $AAB$ configuration. In both cases, the relative size of the non-compensating region decreases monotonically with increasing dilution, providing quantitative confirmation of the disorder-driven stabilization of compensation behaviour.

To characterize the phase boundaries quantitatively, the compensation curves are fitted using a quadratic polynomial form in the interaction-parameter space. For both trilayer geometries, the generic fitting expression is written as

\begin{equation}
	\left[ \dfrac{J_{AA}}{J_{BB}}(\rho) \right]_{\mathrm{fit}}
	=
	p(\rho)
	\left(
	\dfrac{J_{AB}}{J_{BB}}
	\right)^{2}
	+
	q(\rho)
	\left(
	\dfrac{J_{AB}}{J_{BB}}
	\right)
	+
	r(\rho),
\end{equation}

where the fitting coefficients $p(\rho)$, $q(\rho)$, and $r(\rho)$ depend explicitly on the impurity concentration $\rho$. The extracted fitting parameters, together with the corresponding statistical indicators such as $\chi^{2}/n_{\mathrm{dof}}$ and the associated $p$-values, are summarized in Table~\ref{tab_1_phasecurve_fit}. The fitted phase boundaries reproduce the numerically obtained compensation curves with satisfactory accuracy across the investigated dilution range. The systematic variation of the fitting coefficients with impurity concentration further confirms that quenched disorder continuously modifies the topology of the compensation phase space. Consequently, the present results establish that nonmagnetic site dilution provides a viable route for engineering compensation regions and tailoring thermomagnetic response in layered ferrimagnetic heterostructures.

\begin{table}[!htb]	
	\begin{tabular}{|c|c|c|c|c|c|}
		\hline 
		\multicolumn{6}{|c|}{\textbf{(a) ABA configuration}}\\
		\multicolumn{6}{|c|}{$\left[ \dfrac{J_{AA}}{J_{BB}}(\rho) \right]_{\mathrm{fit}} = p_{1}(\rho) \left(\dfrac{J_{AB}}{J_{BB}}\right)^{2} + q_{1}(\rho) \left(\dfrac{J_{AB}}{J_{BB}}\right) + r_{1}(\rho)$} \\ 
		\hline 
		$\rho$ & $p_{1}(\rho)$ & $q_{1}(\rho)$ & $r_{1}(\rho)$ & $\chi^{2}/n_{\mathrm{dof}}$ & $p$-value \\ 
		\hline 
		0.00 & 0.255 & 0.992 & 0.962 & 0.416 & 0.742 \\
		0.05 & 0.158 & 0.943 & 1.035 & 0.338 & 0.852 \\
		0.10 & 0.072 & 0.846 & 1.087 & 0.087 & 0.967 \\
		0.15 & 0.145 & 0.871 & 1.156 & 0.044 & 0.988 \\
		0.20 & 0.318 & 1.067 & 1.295 & 0.308 & 0.735 \\
		0.25 & 0.686 & 1.806 & 1.781 & -- & -- \\
		0.30 & 0.234 & 0.948 & 1.531 & -- & -- \\
		\hline
	\end{tabular} 
	
	\vskip 0.3cm
	
	\begin{tabular}{|c|c|c|c|c|c|}
	\hline 
	\multicolumn{6}{|c|}{\textbf{(b) AAB configuration}}\\
	\multicolumn{6}{|c|}{$\left[ \dfrac{J_{AA}}{J_{BB}}(\rho) \right]_{\mathrm{fit}} = p_{2}(\rho) \left(\dfrac{J_{AB}}{J_{BB}}\right)^{2} + q_{2}(\rho) \left(\dfrac{J_{AB}}{J_{BB}}\right) + r_{2}(\rho)$} \\ 
	\hline 
	$\rho$ & $p_{2}(\rho)$ & $q_{2}(\rho)$ & $r_{2}(\rho)$ & $\chi^{2}/n_{\mathrm{dof}}$ & $p$-value \\ 
	\hline 
	0.00 & 0.167 & 0.390 & 0.701 & 0.155 & 0.961 \\
	0.05 & 0.079 & 0.312 & 0.746 & 0.063 & 0.993 \\
	0.10 & 0.113 & 0.334 & 0.810 & 0.264 & 0.901 \\
	0.15 & 0.144 & 0.382 & 0.894 & 0.061 & 0.993 \\
	0.20 & 0.252 & 0.503 & 0.987 & 0.025 & 0.999 \\
	0.25 & 0.118 & 0.321 & 1.043 & 0.117 & 0.950 \\
	0.30 & 0.019 & 0.137 & 1.049 & 0.290 & 0.748 \\
	\hline
\end{tabular} 

	\caption{Fitting parameters associated with the compensation phase boundaries for the diluted $ABA$ and $AAB$ Ising trilayer ferrimagnetic configurations.}
	\label{tab_1_phasecurve_fit}
\end{table}

To further quantify the disorder-driven evolution of the compensation phase space, the impurity dependence of the fractional non-compensating area was analysed using simple polynomial fitting functions. For the $ABA$ trilayer configuration, the variation of the normalized non-compensating area is accurately described by a linear relation of the form:
\begin{equation}
f_{1}(\rho)
=
\left[
\dfrac{A(\rho)}{A_{\mathrm{tot}}}
\right]_{ABA}
=
n_{1}\rho + l_{1}.
\end{equation}

Extrapolation of the fitted curve yields a critical impurity concentration
\[
\rho_{c,ABA}=0.3186\pm0.0114,
\]
beyond which the non-compensating phase region disappears entirely within the investigated interaction domain. Consequently, for $\rho>\rho_{c,ABA}$, all explored combinations of reduced exchange interaction strengths exhibit compensation behaviour.

In contrast, the corresponding behaviour for the $AAB$ trilayer geometry is better described by a quadratic polynomial of the form:
\begin{equation}
f_{2}(\rho)
=
\left[
\dfrac{A(\rho)}{A_{\mathrm{tot}}}
\right]_{AAB}
=
m_{2}\rho^{2}+n_{2}\rho+l_{2}.
\end{equation}

The resulting fit provides an estimated critical impurity concentration
\[
\rho_{c,AAB}=0.3126\pm0.0462.
\]

Similar to the $ABA$ geometry, impurity concentrations exceeding $\rho_{c,AAB}$ lead to complete suppression of the non-compensating region in the explored parameter space, implying that all investigated interaction-parameter combinations display compensation phenomena.

These results establish the existence of an approximate threshold concentration of quenched nonmagnetic impurities for both trilayer geometries. Above this threshold, compensation behaviour becomes universally stabilized throughout the investigated Hamiltonian parameter range. The emergence of such disorder-controlled critical concentrations further demonstrates that site dilution can serve as an efficient mechanism for engineering compensation phenomena and tailoring thermomagnetic functionality in layered Ising ferrimagnetic heterostructures.

\subsection{Systematics}

\begin{figure*}[!htb]
	\begin{center}
		\begin{tabular}{c}
			
			\resizebox{8.5cm}{!}{\includegraphics[angle=0]{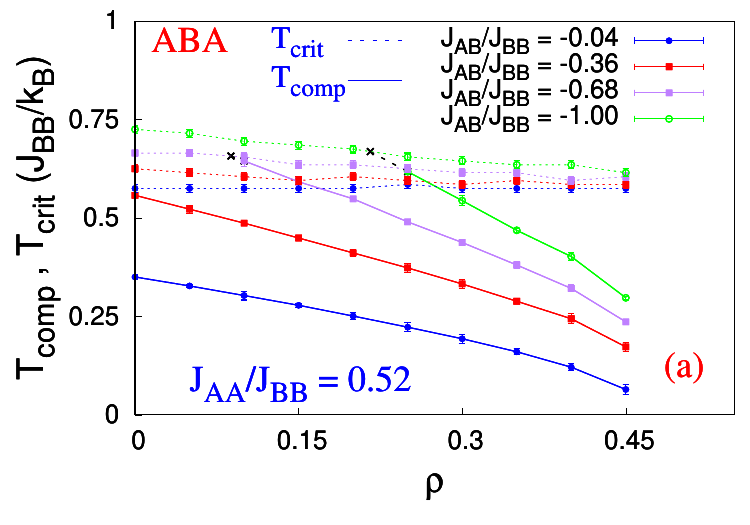}}
			\resizebox{8.5cm}{!}{\includegraphics[angle=0]{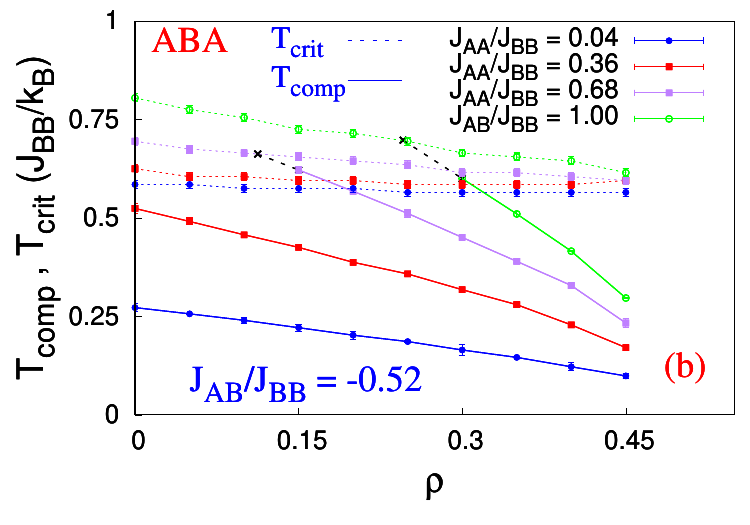}}\\
			
			\resizebox{8.5cm}{!}{\includegraphics[angle=0]{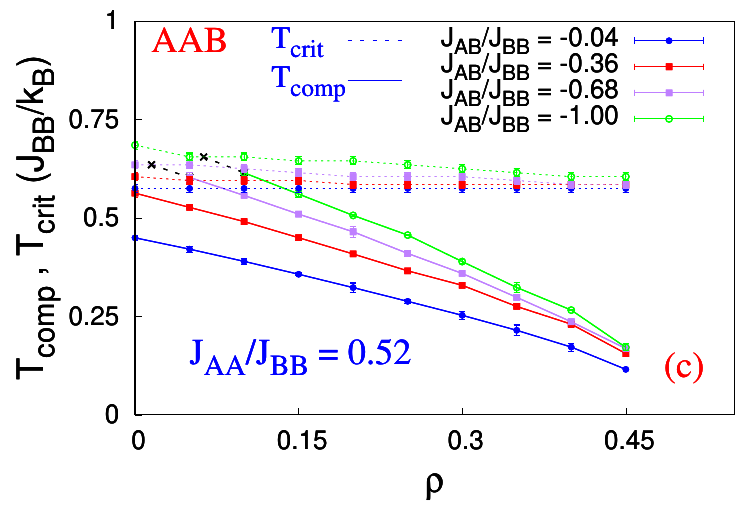}}
			\resizebox{8.5cm}{!}{\includegraphics[angle=0]{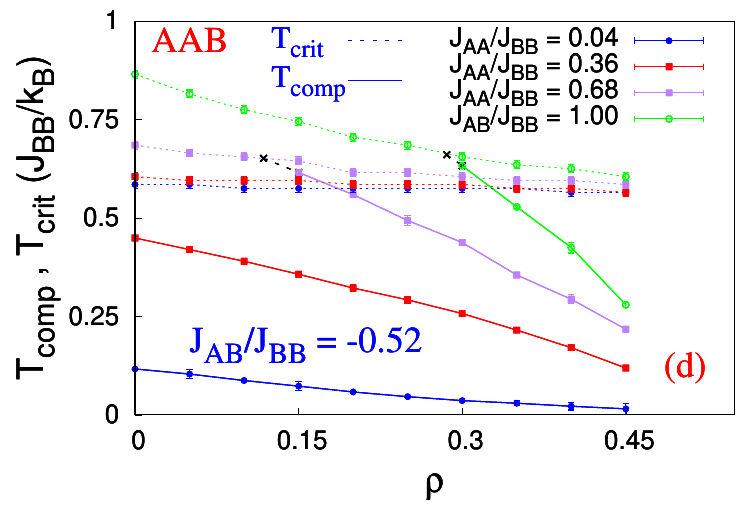}}
			
		\end{tabular}
		\caption{(Colour Online) Variation of the compensation temperature $T_{\mathrm{comp}}$ and critical temperature $T_{c}$ with increasing concentration of nonmagnetic impurities $\rho$ for representative coupling combinations: (a,b) $ABA$ trilayer configuration and (c,d) $AAB$ trilayer configuration. Error bars smaller than the point markers are not visible.}
	\label{fig_7_tcomp_tcrit_rho}
	\end{center}
\end{figure*}

We now investigate the systematic evolution of the compensation and critical temperatures with increasing concentration of quenched nonmagnetic impurities for representative combinations of coupling strengths. Figure~\ref{fig_7_tcomp_tcrit_rho} summarizes the behaviour for both the $ABA$ and $AAB$ diluted Ising trilayer configurations.

For both geometries, the compensation temperature exhibits a distinctly nonlinear dependence on the impurity concentration $\rho$ up to a characteristic threshold concentration. Beyond this threshold value, the thermomagnetic response changes qualitatively and compensation becomes stabilized over the entire investigated interaction regime. In this context, the threshold concentration of nonmagnetic impurities may be defined as the minimum concentration of spin-$0$ atoms above which the compensation temperature and the critical temperature become clearly distinguishable for a given set of coupling parameters.

The behaviour of the critical temperature differs substantially from that of the compensation temperature. Even for relatively strong dilution, the suppression of the critical temperature remains comparatively moderate for weak interaction strengths. For the weakest interaction combination investigated, namely $J_{AA}/J_{BB}=0.04$ and $J_{AB}/J_{BB}=-0.04$, the variation of the critical temperature remains within approximately $1.74\%$ up to $45\%$ dilution for both trilayer configurations. In contrast, for the strongest interaction regime considered, $J_{AA}/J_{BB}=1.00$ and $J_{AB}/J_{BB}=-1.00$, the reduction in the critical temperature reaches nearly $27.32\%$ for the $ABA$ configuration and approximately $29.51\%$ for the $AAB$ configuration. These observations indicate that quenched disorder influences compensation behaviour much more strongly than the overall magnetic ordering temperature, particularly within the weak-coupling regime.

\begin{figure*}[!htb]
	\begin{center}
		\begin{tabular}{c}
			
			\resizebox{8.5cm}{!}{\includegraphics[angle=0]{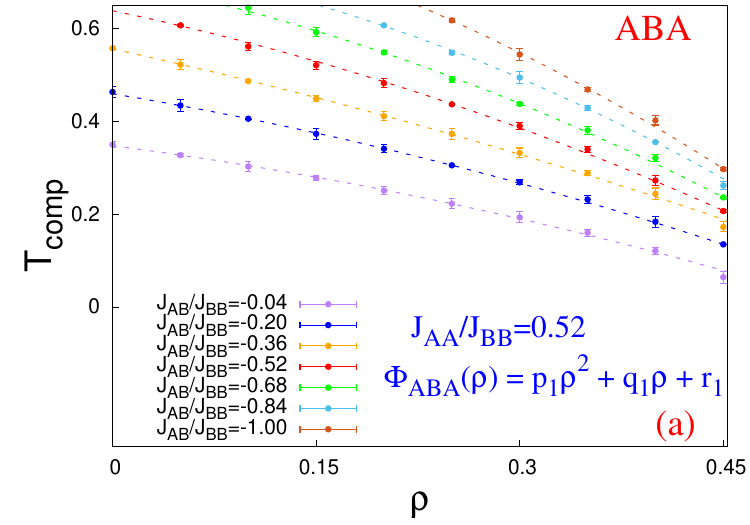}}
			\resizebox{8.5cm}{!}{\includegraphics[angle=0]{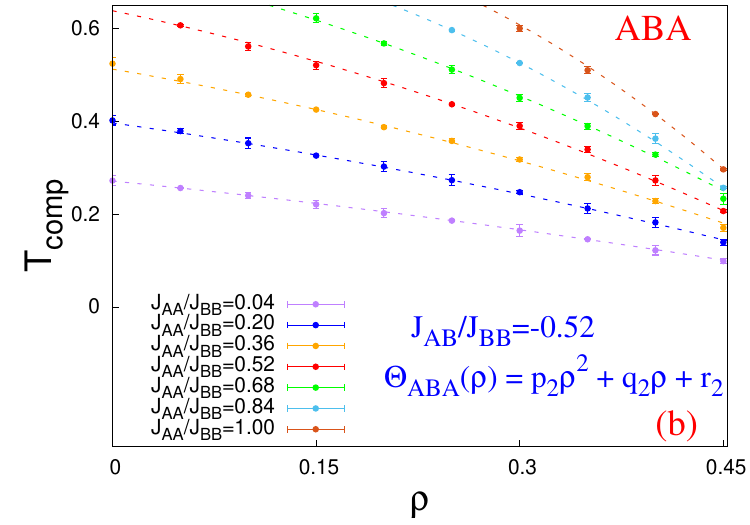}}\\
			
			\resizebox{8.5cm}{!}{\includegraphics[angle=0]{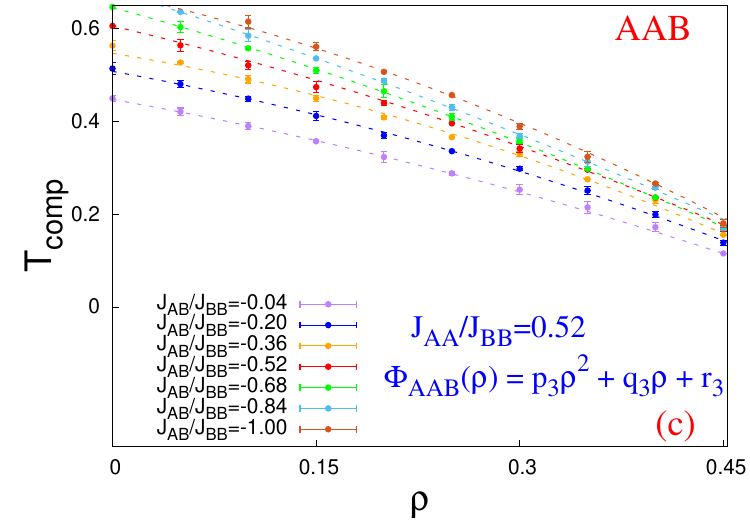}}
			\resizebox{8.5cm}{!}{\includegraphics[angle=0]{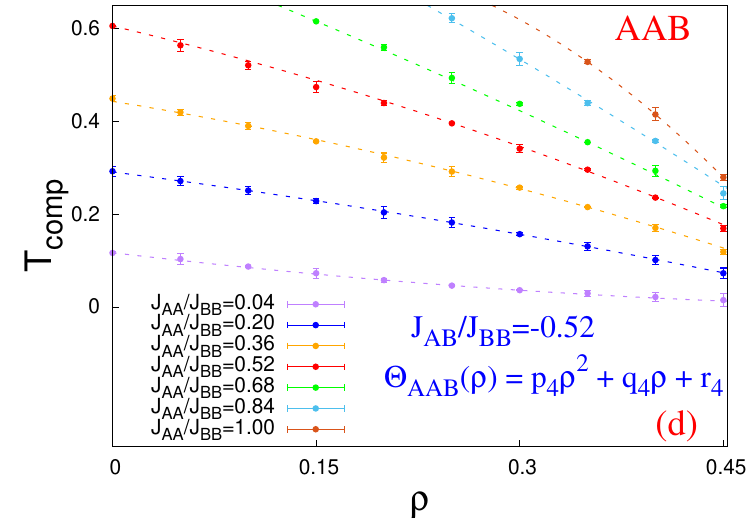}}
			
		\end{tabular}
	\caption{(Colour Online) Quadratic fitting behaviour for the variation of the compensation temperature with impurity concentration $\rho$ for representative interaction strengths: (a,b) $ABA$ trilayer configuration and (c,d) $AAB$ trilayer configuration. Error bars smaller than the point markers are not visible.}
	\label{fig_8_tcomp_rho_fit}
	\end{center}
\end{figure*}

A natural question emerging from the preceding observations concerns the possibility of constructing an effective phenomenological description for the variation of the compensation temperature with impurity concentration. Figure~\ref{fig_8_tcomp_rho_fit} demonstrates that a quadratic polynomial provides an excellent approximation to the simulation data throughout the investigated parameter range.

For the $ABA$ trilayer configuration, the compensation temperature is fitted using
\begin{equation}
\label{eq_tcomp_fit_ABA}
T_{\mathrm{comp},ABA}
\left(
\rho,
\frac{J_{AA}}{J_{BB}},
\frac{J_{AB}}{J_{BB}}
\right)
=
p_{1}\rho^{2}+q_{1}\rho+r_{1},
\end{equation}
whereas for the $AAB$ trilayer configuration we employ
\begin{equation}
\label{eq_tcomp_fit_AAB}
T_{\mathrm{comp},AAB}
\left(
\rho,
\frac{J_{AA}}{J_{BB}},
\frac{J_{AB}}{J_{BB}}
\right)
=
p_{2}\rho^{2}+q_{2}\rho+r_{2}.
\end{equation}

The fitting coefficients
$p_{i}\equiv p_{i}(J_{AA}/J_{BB},J_{AB}/J_{BB})$,
$q_{i}\equiv q_{i}(J_{AA}/J_{BB},J_{AB}/J_{BB})$,
and
$r_{i}\equiv r_{i}(J_{AA}/J_{BB},J_{AB}/J_{BB})$
with $i\in\{1,2\}$ explicitly depend on the reduced coupling strengths and therefore encode the interaction-dependent response of the trilayer system to quenched disorder. Details regarding the fitting procedure and the associated statistical analysis are provided in Appendix~\ref{appendix_fit} \cite{Press}.

Having established the disorder dependence of the compensation temperature, we next investigate the scaling behaviour of the non-compensating region in the Hamiltonian parameter space. In particular, we analyse how the phase area without compensation evolves with the concentration of nonmagnetic impurities. Motivated by the numerical behaviour of the phase diagrams, we propose the following phenomenological scaling relation:
\begin{equation}
\ln\left|\frac{A(\rho)}{\tilde{A}}\right| = ae^{b\rho},
\label{eq_phase_scaling}
\end{equation}
where $A(\rho)$ denotes the non-compensating phase area at impurity concentration $\rho$, while $a$, $b$, and $\tilde{A}$ are effective scaling parameters that depend on the trilayer geometry.

\begin{figure}[!htb]
		\resizebox{9.0cm}{!}{\includegraphics[angle=0]{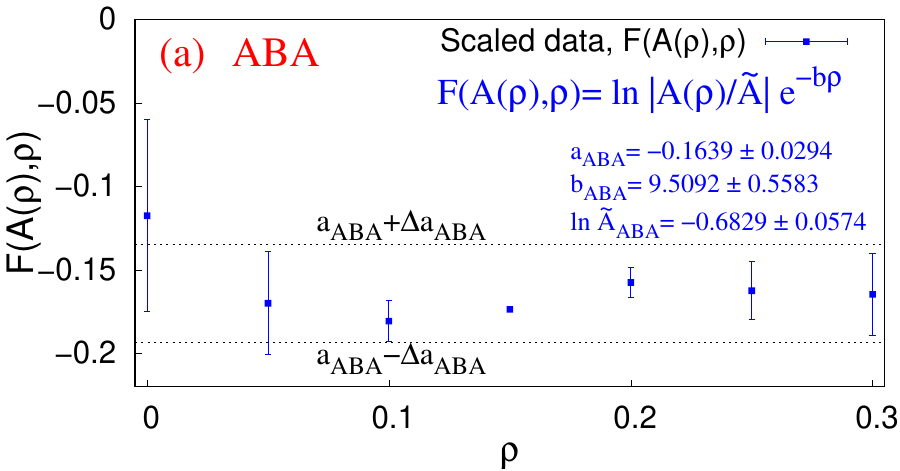}}\\
		
		\resizebox{9.0cm}{!}{\includegraphics[angle=0]{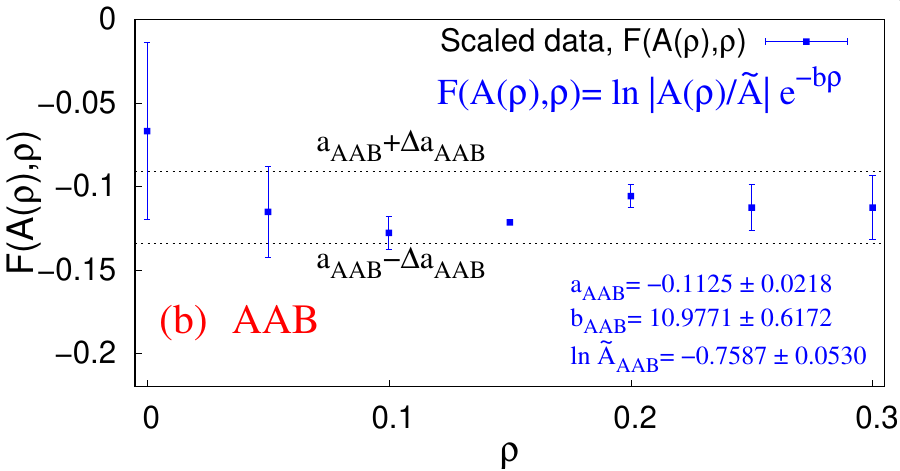}}\\
			
		\caption{(Colour Online) Verification of the proposed scaling behaviour for diluted Ising trilayers: (a) $ABA$ configuration and (b) $AAB$ configuration.} 
	
		\label{fig_9_phase_scaling}
\end{figure}

Regression analysis shows that the proposed scaling form reproduces the numerical data remarkably well for both the $ABA$ and $AAB$ configurations. The extracted fitting parameters are summarized in Table~\ref{tab_2_param_scaling}. Figure~\ref{fig_9_phase_scaling} further demonstrates that the scaled data remain well confined within the fitted interval, thereby supporting the validity of the proposed scaling hypothesis.

\begin{table}[!htb]	
	\resizebox{\columnwidth}{!}{\begin{tabular}{|c|c|c|c|c|c|}
		\hline 
		\multicolumn{6}{|c|}{$\ln\left|\dfrac{A(\rho)}{\tilde{A}}\right| = ae^{b\rho}$} \\ 
		\hline 
		$type$ & $a$ & $b$ & $\ln\tilde{A}$ & $n_{dof}$ & $\chi^{2}/n_{dof}$ \\ 
		\hline 
		ABA & -0.1639 & 9.5092 & -0.6829 & 4 & 0.0020 \\
		
		& $\pm$ 0.0294 & $\pm$ 0.5583 & $\pm$ 0.0574 &  &  \\
		\hline
		
		AAB & -0.1125 & 10.9771 & -0.7587 & 4 & 0.0025 \\
		
		& $\pm$ 0.0218 & $\pm$ 0.6172 & $\pm$ 0.0530 &  &  \\
		\hline
	\end{tabular}}
	
	\caption{Fitting parameters associated with the scaling relation given in Eq.~(\ref{eq_phase_scaling}) for the diluted $ABA$ and $AAB$ trilayer configurations.}
	\label{tab_2_param_scaling}
\end{table}

We now return to the concept of the threshold impurity concentration $\rho_{T}$, defined as the minimum concentration above which compensation behaviour appears for all investigated combinations of coupling strengths. Since compensation originates from the competition between intralayer ferromagnetic interactions and interlayer antiferromagnetic coupling, the threshold concentration naturally depends on both reduced Hamiltonian parameters $J_{AA}/J_{BB}$ and $J_{AB}/J_{BB}$.

\begin{figure*}[!htb]
	\begin{center}
		\begin{tabular}{c}
			
			\resizebox{9.0cm}{!}{\includegraphics[angle=0]{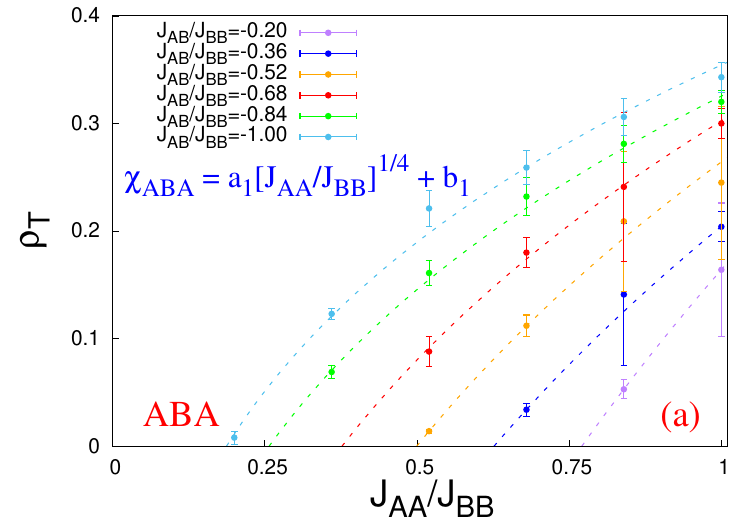}}
			\resizebox{9.0cm}{!}{\includegraphics[angle=0]{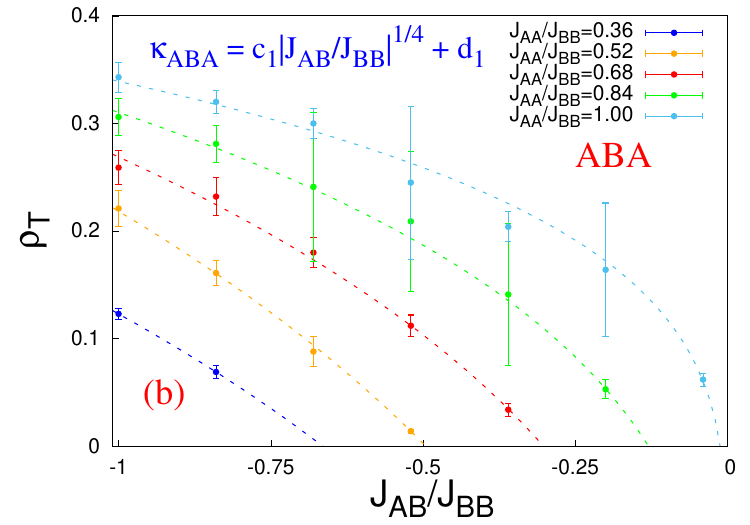}}\\
			
			\resizebox{9.0cm}{!}{\includegraphics[angle=0]{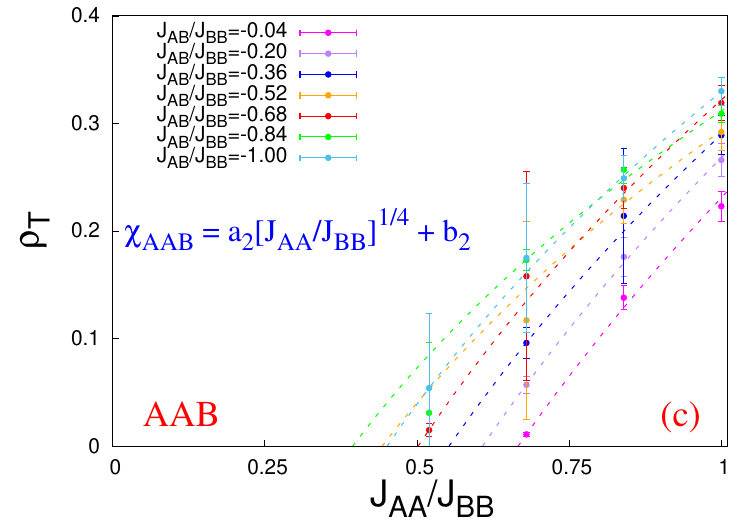}}
			\resizebox{9.0cm}{!}{\includegraphics[angle=0]{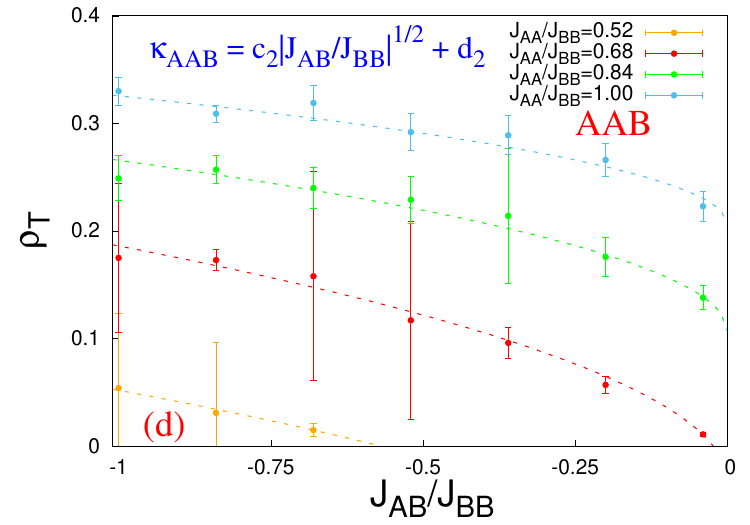}}
			
		\end{tabular}
		\caption{(Colour Online) Threshold impurity concentration $\rho_{T}$ as a function of the reduced coupling strengths for diluted Ising trilayers: (a,b) $ABA$ configuration and (c,d) $AAB$ configuration.}
		\label{fig_10_thres_conc_2d}
	\end{center}
\end{figure*}

Figure~\ref{fig_10_thres_conc_2d} illustrates the systematic variation of $\rho_{T}$ with the reduced coupling strengths for both trilayer geometries. In general, stronger intralayer ferromagnetic interactions or stronger interlayer antiferromagnetic couplings require larger impurity concentrations before universal compensation behaviour emerges. Thus, increasing the magnitude of either coupling ratio systematically shifts the threshold concentration toward higher values.

To model the dependence of $\rho_{T}$ on the reduced coupling strengths, we propose separate phenomenological forms associated with the intralayer and interlayer interaction sectors. For the $ABA$ trilayer configuration, the threshold concentration is described by
\begin{eqnarray}
\left[
\chi_{ABA}
\left(
\dfrac{J_{AA}}{J_{BB}}
\right)
\right]_{\dfrac{J_{AB}}{J_{BB}}}
&=&
a_{1}
\sqrt[4]{\dfrac{J_{AA}}{J_{BB}}}
+b_{1},
\\
\left[
\kappa_{ABA}
\left(
\dfrac{J_{AB}}{J_{BB}}
\right)
\right]_{\dfrac{J_{AA}}{J_{BB}}}
&=&
c_{1}
\sqrt[4]{\left|
\dfrac{J_{AB}}{J_{BB}}
\right|}
+d_{1}.
\end{eqnarray}

Here, the coefficients $a_{1}$ and $b_{1}$ depend on the interlayer coupling ratio, while $c_{1}$ and $d_{1}$ depend on the intralayer interaction strength. Similarly, for the $AAB$ trilayer configuration, we employ
\begin{eqnarray}
\left[
\chi_{AAB}
\left(
\dfrac{J_{AA}}{J_{BB}}
\right)
\right]_{\dfrac{J_{AB}}{J_{BB}}}
&=&
a_{2}
\sqrt[4]{\dfrac{J_{AA}}{J_{BB}}}
+b_{2},
\\
\left[
\kappa_{AAB}
\left(
\dfrac{J_{AB}}{J_{BB}}
\right)
\right]_{\dfrac{J_{AA}}{J_{BB}}}
&=&
c_{2}
\sqrt{
\left|
\dfrac{J_{AB}}{J_{BB}}
\right|
}
+d_{2}.
\end{eqnarray}

As before, the coefficients $a_{2}$ and $b_{2}$ depend on the interlayer interaction strength, whereas $c_{2}$ and $d_{2}$ depend on the intralayer coupling ratio. An effective estimate of the threshold concentration may then be obtained phenomenologically through the geometric mean of the two contributions:
\begin{equation}
\rho_{T}
\left(
\dfrac{J_{AA}}{J_{BB}},
\dfrac{J_{AB}}{J_{BB}}
\right)
=
\sqrt{
\left[
\chi
\left(
\dfrac{J_{AA}}{J_{BB}}
\right)
\right]_{\dfrac{J_{AB}}{J_{BB}}}
\left[
\kappa
\left(
\dfrac{J_{AB}}{J_{BB}}
\right)
\right]_{\dfrac{J_{AA}}{J_{BB}}}
}.
\end{equation}

Although phenomenological in nature, these functional forms successfully capture the systematic disorder-driven evolution of the compensation landscape and provide an effective framework for estimating the impurity concentration required to induce compensation in diluted Ising trilayer ferrimagnets.

\section{Summary}
\label{sec_summary}

In the present work, we have systematically investigated the influence of quenched site dilution on the thermodynamic response and compensation phenomena in ferrimagnetic spin-$1/2$ Ising trilayers composed of square monolayers. Two distinct stacking geometries, namely the $ABA$ and $AAB$ configurations, were considered, where only the A-type layers are subjected to random site dilution while the strongly coupled B-layer remains magnetically pristine. Using extensive Metropolis Monte Carlo simulations, we analysed how the interplay between intra-layer ferromagnetic interactions, inter-layer antiferromagnetic interactions, and non-magnetic impurity concentration governs the emergence and evolution of compensation behaviour in these non-equivalent trilayered systems.

The numerical analysis demonstrates that quenched disorder strongly modifies the thermomagnetic response of both trilayer configurations. For fixed impurity concentration, increasing the magnitudes of the reduced coupling strengths shifts both the compensation and critical temperatures toward higher values, while the compensation point progressively approaches the critical temperature before eventually disappearing. Conversely, for fixed coupling ratios, increasing the concentration of non-magnetic impurities suppresses both the compensation and transition temperatures. The evolution of the magnetisation profiles further reveals a sequence of impurity-driven transformations between distinct Néel classifications, including transitions from non-compensating Q- and R-type behaviours to compensating N-type profiles. These transformations are consistently accompanied by characteristic signatures in the magnetic susceptibility and specific heat curves, where broadened anomalies emerge near the compensation temperature in addition to the critical singularities associated with the magnetic phase transition.

One of the central outcomes of the present study is the observation that quenched site dilution can induce compensation behaviour in parameter regimes where the pristine systems exhibit no compensation point. In both $ABA$ and $AAB$ trilayers, the non-compensating region in the Hamiltonian parameter space systematically shrinks with increasing impurity concentration. Beyond a threshold concentration of non-magnetic atoms, all investigated combinations of coupling strengths exhibit compensation behaviour. This establishes site dilution as an effective mechanism for engineering compensation phenomena in layered ferrimagnetic systems.

The compensation phase boundaries in the parameter space $\left(J_{AB}/J_{BB}\times J_{AA}/J_{BB}\right)$ were quantitatively characterised through polynomial fitting schemes, while Monte Carlo integration was employed to estimate the fractional area of the non-compensating region. The resulting analysis revealed well-defined scaling behaviour of the phase area with impurity concentration. Furthermore, the compensation temperature itself was shown to exhibit a robust nonlinear dependence on dilution and can be accurately described by quadratic fitting functions whose coefficients depend explicitly on the reduced coupling strengths. The threshold impurity concentration required to generate compensation behaviour was also found to vary systematically with both ferromagnetic and antiferromagnetic coupling ratios, allowing the formulation of semi-empirical functional relations capable of predicting the onset of compensation.

Overall, the present investigation establishes a comprehensive thermodynamic and systematic framework for understanding impurity-driven compensation phenomena in ferrimagnetic Ising trilayers with non-equivalent magnetic sublattices. Beyond providing detailed numerical characterization of disorder-induced effects, the proposed scaling relations and phenomenological modelling offer predictive insight into the manipulation of compensation behaviour through controlled dilution. Since compensation phenomena are of considerable technological relevance in thermomagnetic switching, spintronic devices, and magneto-optical recording applications, the present results may provide useful theoretical guidance for the design and optimization of artificially engineered layered ferrimagnetic materials with tunable magnetic compensation properties.

\section*{Acknowledgements}
The author gratefully acknowledges financial assistance from the University Grants Commission, India in the form of a Research fellowship and extends his thanks to Dr. Tamaghna Maitra, Ms. Esha Lyngdoh and Dr. W L Reenbohn for providing the computational facilities and to Dr. Soumyajit Sarkar and Dr. Rupak Kumar Bhattacharya for their insightful comments.

\vskip 1.0 cm
\bibliography{myref}   
\bibliographystyle{unsrt}

\end{document}